\documentclass[fleqn,usenatbib]{mnras}

\usepackage{newtxtext,newtxmath}

\usepackage[T1]{fontenc}
\usepackage{ae,aecompl}

\usepackage{graphicx}	
\usepackage{amsmath}	
\usepackage{amssymb}	
\usepackage{subfigure}

\newcommand{\kms}{\ifmmode{\,\rm{km}\, \rm{s}^{-1}}\else{$\,$km$\,$s$^{-1}$}\fi}
\newcommand{\msun}{\ifmmode{M_{\odot}}\else{$M_{\odot}$}\fi}
\newcommand{\mstar}{\ifmmode{M_{\star}}\else{$M_{\star}$}\fi}

\newcommand{\rmaxs}{\ifmmode{R_{\rm{\sigma}}^{\rm{max}}}\else{$R_{\rm{\sigma}}^{\rm{max}}$}\fi}
\newcommand{\recirc}{\ifmmode{R_{\rm{e,c}}}\else{$R_{\rm{e,c}}$}\fi}
\newcommand{\re}{\ifmmode{R_{\rm{e}}}\else{$R_{\rm{e}}$}\fi}
\newcommand{\ret}{\ifmmode{R_{\rm{e/2}}}\else{$R_{\rm{e/2}}$}\fi}
\newcommand{\retwo}{\ifmmode{R_{\rm{e/2}}}\else{$2R_{\rm{e}}$}\fi}
\newcommand{\aee}{\ifmmode{a_{\rm{e}}}\else{$a_{\rm{e}}$}\fi}
\newcommand{\kpa}{\ifmmode{PA_{\rm{kin}}}\else{$PA_{\rm{kin}}$}\fi}
\newcommand{\ee}{\ifmmode{\epsilon_{\rm{e}}}\else{$\epsilon_{\rm{e}}$}\fi}
\newcommand{\lr}{\ifmmode{\lambda_R}\else{$\lambda_{R}$}\fi}
\newcommand{\lre}{\ifmmode{\lambda_{R_{\rm{e}}}}\else{$\lambda_{R_{\rm{e}}}$}\fi}
\newcommand{\lret}{\ifmmode{\lambda_{R_{\rm{e/2}}}}\else{$\lambda_{R_{\rm{e/2}}}$}\fi}
\newcommand{\lretwo}{\ifmmode{\lambda_{2R_{\rm{e}}}}\else{$\lambda_{2R_{\rm{e}}}$}\fi}
\newcommand{\flr}{\ifmmode{f_{\lambda_{R}}}\else{$f_{\lambda_{R_{\rm{e}}}}$}\fi}
\newcommand{\flre}{\ifmmode{f_{\lambda_{R_{\rm{e}}}}}\else{$f_{\lambda_{R_{\rm{e}}}}$}\fi}
\newcommand{\flret}{\ifmmode{f_{\lambda_{R_{\rm{e/2}}}}}\else{$f_{\lambda_{R_{\rm{e/2}}}}$}\fi}
\newcommand{\vs}{\ifmmode{V / \sigma}\else{$V / \sigma$}\fi}
\newcommand{\vse}{\ifmmode{(V / \sigma)_{\rm{e}}}\else{$(V / \sigma)_{\rm{e}}$}\fi}
\newcommand{\vset}{\ifmmode{(V / \sigma)_{\rm{e/2}}}\else{$(V / \sigma)_{\rm{e/2}}$}\fi}
\newcommand{\vsetwo}{\ifmmode{(V / \sigma)_{\rm{2e}}}\else{$(V / \sigma)_{\rm{2e}}$}\fi}
\newcommand{\vobs}{\ifmmode{V_{\rm{obs}}}\else{$V_{\rm{obs}}$}\fi}
\newcommand{\sobs}{\ifmmode{\sigma_{\rm{obs}}}\else{$\sigma_{\rm{obs}}$}\fi}
\newcommand{\kinemetry}{\textsc{kinemetry}}

\title[SAMI Galaxy Survey: Data Release Two]{The SAMI Galaxy Survey: Data Release Two with absorption-line physics value-added products}

\author[N. Scott et al.]{
Nicholas Scott$^{1,2,3}$\thanks{E-mail: nicholas.scott@sydney.edu.au (NS)},
Jesse van de Sande$^{1,3}\thanks{E-mail: jesse.vandesande@sydney.edu.au (JvdS)}$,
Scott M. Croom$^{1,2,3}$,\newauthor
Brent Groves$^{2,3,4}$,
Matt S. Owers$^{5,6}$,
Henry Poetrodjojo$^{2,3,4}$,
Francesco D'Eugenio$^{2,4}$,\newauthor
Anne M. Medling$^{4,7}$\thanks{Hubble Fellow},
Dilyar Barat$^{2,3,4}$,
Tania M. Barone$^{1,2,3,4}$,\newauthor
Joss Bland-Hawthorn$^{1,3}$,
Sarah Brough$^{2,3,8}$,
Julia Bryant$^{1,2,3,9}$,
Luca Cortese$^{3,10}$,\newauthor
Caroline Foster$^{1,3}$,
Andrew W. Green$^{6}$,
Sree Oh$^{3,4}$,
Matthew Colless$^{2,3,4}$,\newauthor
Michael J. Drinkwater$^{11}$,
Simon P. Driver$^{10}$,
Michael Goodwin$^{6,12}$,\newauthor
Madusha L. P. Gunawardhana$^{13}$,
Christoph Federrath$^{3,4}$,
Lloyd Harischandra$^{6,12}$,\newauthor
Yifei Jin$^{3,4}$,
J. S. Lawrence$^{6,12}$,
Nuria P. Lorente$^{6,12}$,
Elizabeth Mannering$^{6,10,12}$,\newauthor
Simon O'Toole$^{6,12}$,
Samuel N. Richards$^{14}$,
Sebastian F. Sanchez$^{15}$,\newauthor
Adam L. Schaefer$^{1,2,16}$,
Katrina Sealey$^{6,12}$,
Rob Sharp$^{4}$,
Sarah M. Sweet$^{3,17}$\newauthor
Dan S. Taranu$^{2,10}$
and Mathew Varidel$^{1,3}$
\\
Affiliations are listed at the end of the paper
}

\date{Accepted XXX. Received YYY; in original form ZZZ}

\pubyear{2018}

\begin{document}
\label{firstpage}
\pagerange{\pageref{firstpage}--\pageref{lastpage}}
\maketitle

\begin{abstract}
We present the second major release of data from the SAMI Galaxy Survey. Data Release Two includes data for 1559 galaxies, about 50\% of the full survey. Galaxies included have a redshift range $0.004 < z < 0.113$ and a large stellar mass range $7.5 < \log (M_\star/M_\odot) < 11.6$. The core data for each galaxy consist of two primary spectral cubes covering the blue and red optical wavelength ranges. For each primary cube we also provide three spatially binned spectral cubes and a set of standardised aperture spectra. For each core data product we provide a set of value-added data products. This includes all emission line value-added products from Data Release One, expanded to the larger sample.  In addition we include stellar kinematic and stellar population value-added products derived from absorption line measurements. The data are provided online through Australian Astronomical Optics' Data Central. We illustrate the potential of this release by presenting the distribution of $\sim 350,000$ stellar velocity dispersion measurements from individual spaxels as a function of $R/\re$, divided in four galaxy mass bins. In the highest stellar mass bin ($\log (M_\star/M_\odot)>11$), the velocity dispersion strongly increases towards the centre, whereas below  $\log (M_\star/M_\odot)<10$ we find no evidence for a clear increase in the central velocity dispersion. This suggests a transition mass around $\log (M_\star/M_\odot)~\sim10$ for galaxies with or without a dispersion--dominated bulge.

\end{abstract}

\begin{keywords}
galaxies: general - galaxies: kinematics and dynamics - galaxies: abundances - galaxies: star formation - galaxies: stellar content - astronomical data bases: surveys
\end{keywords}



\section{Introduction}

Galaxies are composed of multiple distinct components, such as thin and thick disk, bulge, bar(s), spiral arms, ring(s) and many others. As well as having different spatial structure, these components can also differ in terms of their kinematics and their compositions; either narrowly in terms of differing chemistry, or more broadly into stellar, gaseous and dark matter. Determining how these distinct components interact and change over time is critical to a deeper understanding of galaxy evolution \citep[e.g.,][]{mo2010}.

Spatially resolving galaxies is essential to understand the different components. While imaging studies, particularly multi-wavelength imaging, can begin to disentangle these components, access to kinematic and chemical separation is largely unavailable. Spatially resolved spectroscopy is ideally suited to this task, as the simultaneous separation of the observed light, both spectrally and spatially, provides the most detailed dissection of the internal structure of galaxies currently available \citep[e.g.,][]{cappellari2016}.

The challenge of spatially resolved spectroscopy is that spreading the light out both spatially and spectrally drastically reduces the signal-to-noise ratio (S/N) per resolving element. This limitation has restricted initial work in this area to relatively small samples of objects, or to specific classes of object that are more easily observed. While these kinds of studies have been very successful in addressing the role of specific physical processes that shape galaxies, a broader view is required to develop a holistic understanding of galaxy evolution.

Galaxies are very diverse, and the physical processes involved in galaxy evolution are many and varied. To fully understand the primary drivers of galaxy evolution one requires large samples that encompass the complete diversity of the galaxy population. This can be achieved with spatially resolved spectroscopy by either investing large amounts of telescope time with a single-object instrument, e.g. surveys with $N>250$, ATLAS$^{\rm{3D}}$ \citep{cappellari2011a} and CALIFA \citep{sanchez2012}, or through the use of a multiplexed integral field spectrograph, such as SAMI \citep{croom2012}, MaNGA \citep{bundy2015} or KMOS \citep{wisnioski2015,stott2016}. Since the beginning of this decade, large integral field spectroscopy surveys such as these, have been assembling samples of hundreds or thousands of galaxies, allowing us to dissect, in detail, the entire population of local galaxies.

\citet[SAMI Data Release 1, hereafter DR1]{green2018} discussed the broad role integral field spectroscopy has played in furthering our understanding of galaxies. In the current release we hope to push forward the exploration of new analyses that utilise the full power of combining emission, absorption and dynamical measurements by providing extensive value added data products.

In this paper we present the second public release (DR2) of SAMI Galaxy Survey observations, including both fully processed spectral data cubes and a large array of derived science products that enable an extremely broad approach for studying galaxies. In Section \ref{sec:overview} we briefly review the survey and instrument design and progress on observing and processing the data since DR1. In Section \ref{sec:dr2_sample} we describe the galaxies presented in this sample. In Section \ref{sec:core_data} we describe the core data of this release; spatially resolved spectral cubes and additional spectral data derived from these cubes, and the quality of the data. In Section \ref{sec:emission_line_products} we describe the emission line products included in this release, with a focus on changes since DR1, and in Section \ref{sec:absorption_line_products} we describe the new absorption line products being released for the first time. Finally, in Section \ref{sec:data_access} we describe how this data can be accessed through the Data Central web service and provide an example science use of these data to illustrate the potential power of this data release. Throughout this release we adopt the concordance cosmology: ($\Omega_\Lambda, \Omega_m, h) = (0.7, 0.3, 0.7)\ $\citep{hinshaw2009}.

\section{The SAMI Galaxy Survey}
\label{sec:overview}
The SAMI Galaxy Survey \citep{bryant2015} is a spatially resolved spectroscopic survey of a large sample of nearby ($z \lesssim 0.1$) galaxies, conducted with the Sydney -- Australian Astronomical Observatory Multi-Object Integral Field Spectrograph \citep[SAMI,][]{croom2012}.

The SAMI instrument is a multi-object Integral Field Spectrograph (IFS) mounted at the prime focus of the 3.9m Anglo-Australian Telescope (AAT). SAMI uses 13 fused optical fibre bundles \citep[hexabundles;][]{blandhawthorn2011,bryant2011,bryant2014} that can be deployed across a 1 degree diameter field of view. Each hexabundle consists of 61 closely packed optical fibres, where each fibre has a diameter of 1.6 arcsec, resulting in an integral field unit (IFU) with a diameter of 15 arcsec and a fill factor of 75 per cent; 26 additional fibres provide simultaneous blank sky observations. 

SAMI feeds the AAOmega optical spectrograph \citep{sharp2006}. The SAMI Galaxy Survey makes use of the 580V and 1000R gratings, with a dichroic to split the light at 5700\ \AA\ between the two spectrograph arms. The precise wavelength coverage and spectral resolution of this instrumental set up is given in Table \ref{tab:specres_table}.

\subsection{Survey sample and observing status}
\label{sec:sample}
The selection of the SAMI Galaxy Survey sample is described in detail in \citet{bryant2015}, with further details in \citet{owers2017}. Here we briefly summarise the primary sample and describe the status of secondary targets. 

The SAMI Galaxy Survey sample consists of two separate but complementary samples with matched selection criteria; a SAMI-GAMA sample drawn from the Galaxy And Mass Assembly (GAMA) survey \citep{driver2011} and an additional cluster sample. The SAMI-GAMA sample consists of a series of volume-limited samples, where the stellar mass limit for each sample increases with redshift. Stellar masses are estimated from the rest-frame $i$-band absolute magnitude and $g-i$ colour by using the colour-mass relation following the method of \citet{taylor2011}, assuming a \citet{chabrier2003} stellar initial mass function (IMF) and exponentially declining star formation histories. The SAMI-GAMA sample is drawn from the three $4 \times 12$ degree fields of the initial GAMA-I survey \citep{driver2011}. These regions include galaxies in a range of environments, from isolated up to massive groups, but do not contain any galaxy clusters within the $z \leq 0.1$ SAMI limit. To complete the environmental coverage, the SAMI Galaxy Survey includes an additional cluster sample, drawn from eight $z \leq 0.1$ clusters, described in \citet{owers2017}. The same stellar mass selection limits were applied to the cluster sample as for the main sample. In practice, for the clusters with $z < 0.045$ we target cluster galaxies with $\log(\mstar/\msun) > 9.5$, and for the clusters with $0.045 < z < 0.06$ we target cluster galaxies with $\log(\mstar/\msun) > 10.0$.

In addition, a sample of secondary target galaxies is defined by galaxies with slightly lower stellar mass cuts in each redshift bin, along with high mass ($\log(\mstar/\msun) > 10.9$) galaxies at slightly higher redshift ($0.095<z<0.115$). The secondary targets were observed when a hexabundle could not be allocated to a primary target. This became necessary as the completeness of the survey grew. In the final semester of observations an extra set of ancillary galaxies were needed to occupy all hexabundles. These were primarily drawn from GAMA galaxies that are in pairs or groups with SAMI galaxies but did not meet the stellar mass cuts of the original selection criteria. None of the ancillary targets are included in DR2.  

Survey observations began in March 2013 and were completed in May 2018. There were a total of 250 observing nights, spread over 34 individual observing runs. At the completion of survey observing, just over 3000 total galaxies were observed. The primary sample was observed to a completeness of 80\% and 84\% in the GAMA and cluster regions respectively with 1930 and 724 unique primary targets in those regions. 

\subsection{Data reduction}
The reduction of SAMI data and the production of data cubes is described fully in \citet{allen2015} and \citet{sharp2015}. Here we briefly summarise the process and in the following section describe in detail the changes since the previous release.

SAMI data reduction broadly falls into two phases; the extraction of row stacked spectra (RSS) from raw observations, and the construction of data cubes from the RSS frames. The creation of RSS frames is handled by the {\sc 2dfDR} data reduction package\footnote{https://www.aao.gov.au/science/software/2dfdr}. Cube creation is carried out using the {\sc SAMI Python} package \citep{allen2014}, and the entire process is automated using the `SAMI Manager', part of the {\sc SAMI Python} package.

Initial reduction consists of the standard steps of overscan subtraction, spectral extraction, flat fielding, fibre throughput calibration, wavelength calibration and sky subtraction. These steps are all accomplished with {\sc 2dfDR}, and result in one RSS frame per observation. Each RSS frame contains data for 12 galaxies and a single calibration star for secondary spectrophotometric calibration and telluric correction.

Relative and absolute flux calibration and telluric correction are applied to each RSS frame using the {\sc SAMI Python} package. The flux-calibrated RSS frames are combined into three-dimensional data cubes by resampling onto a regular grid. This combination includes dither registration and differential atmospheric refraction correction and an additional absolute flux calibration step. The result is a three-dimensional (two spatial and one spectral) data cube. Covariance between spaxels is calculated and stored within the cubes in a compressed form  \cite[see][for details]{sharp2015}.
Binned cubes and aperture spectra are also produced at this stage --- see Sections \ref{subsec:binned_cubes} and \ref{subsec:aperture_spectra} for details.

\subsubsection{Changes between DR1 and DR2}
\label{sec:dr_changes}

For this release, we use the {\sc SAMI Python} package snapshot identified as {\sc mercurial} changeset {\sc 17ebc0ff0a1c}, and {\sc 2dfDR} version 6.65. Several aspects of the data reduction have been improved between these software versions and those used for DR1, which we document in detail below. In addition, the {\sc SAMI Python package} has experienced some quality-of-life improvements including: optimisation of computationally-intensive aspects of the package as compiled {\sc C} code, support for {\sc Python 3} compatibility and increased terminal feedback during data reduction. Three main aspects of the data reduction have been improved for this release. They are: extraction of spectra, flat-fielding and wavelength calibration. 

Spectral extraction requires an accurate trace of the fibre locations across the detector.  These traces (that we call a tram-line map) are derived from a calibration frame, and are based on Gaussian profile fits to each fibre in the spatial direction.  This fit is repeated for each CCD column and the resulting fibre locations are then fitted with a smoothly varying function. For DR1 and earlier releases, this tram-line map was determined from dome flat calibrations taken as part of a standard science observation sequence. However, the dome flat frames have relatively low counts below $\sim 4000$\ \AA, resulting in higher uncertainties in the  tram-line maps in the far blue. For DR2 we used twilight sky frames to derive tram line maps, resulting in more accurate traces below $\sim 4000$\ \AA\ with improved spectral extraction and reduced cross-talk between adjacent fibres. Where twilight sky observations are not available for a given field, we use twilights from different fields.  To account for shifts between tram-line maps from different fields (and on occasion different nights) we measure a 1D (in the spatial direction on the CCD) cross-correlation between the image frames used to generate the tram-line map and the object frame to be extracted.  This also corrects for the small shift caused by the boiling-off of liquid nitrogen in the dewars attached to AAOmega's cameras \citep{sharp2015}.  The cross-correlation is done in $16\times16$ blocks across the CCD and then averaged (with outlier rejection).  This approach allows us to estimate uncertainty on the measured tram-line shift, which is typically a few thousandths of a pixel (standard error on the mean).  Once small bulk shifts in the fibre positions are taken into account we find no difference to final data quality when using a tram-line map derived from a different field. In addition, we have improved the preliminary scattered light model applied before fitting the fibre profiles, which again results in improved spectral extraction, particularly at bluer wavelengths where scattered light represents a larger fraction of the total counts.

\begin{figure*}
\includegraphics[width=0.3\linewidth]{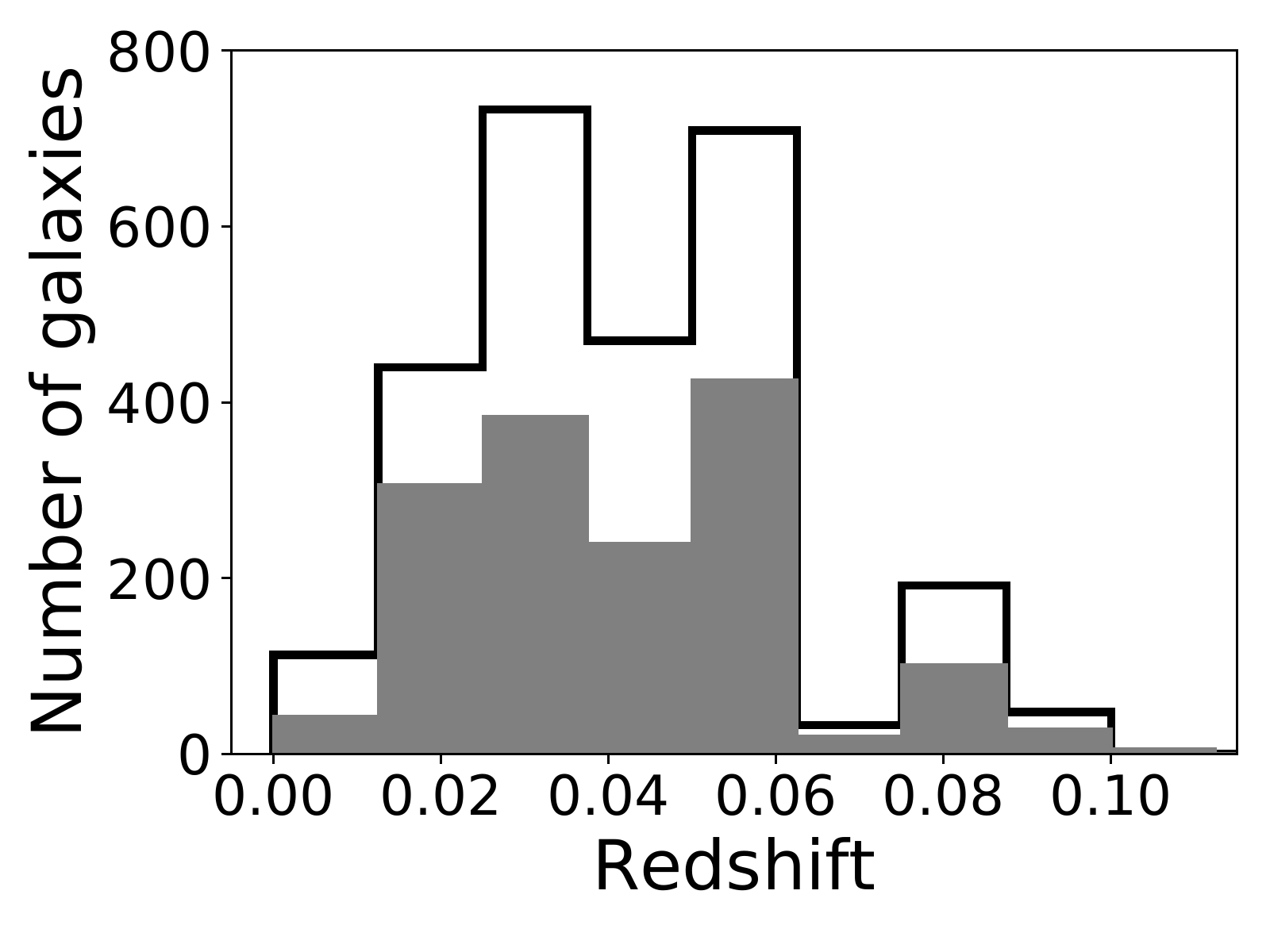}
\includegraphics[width=0.3\linewidth]{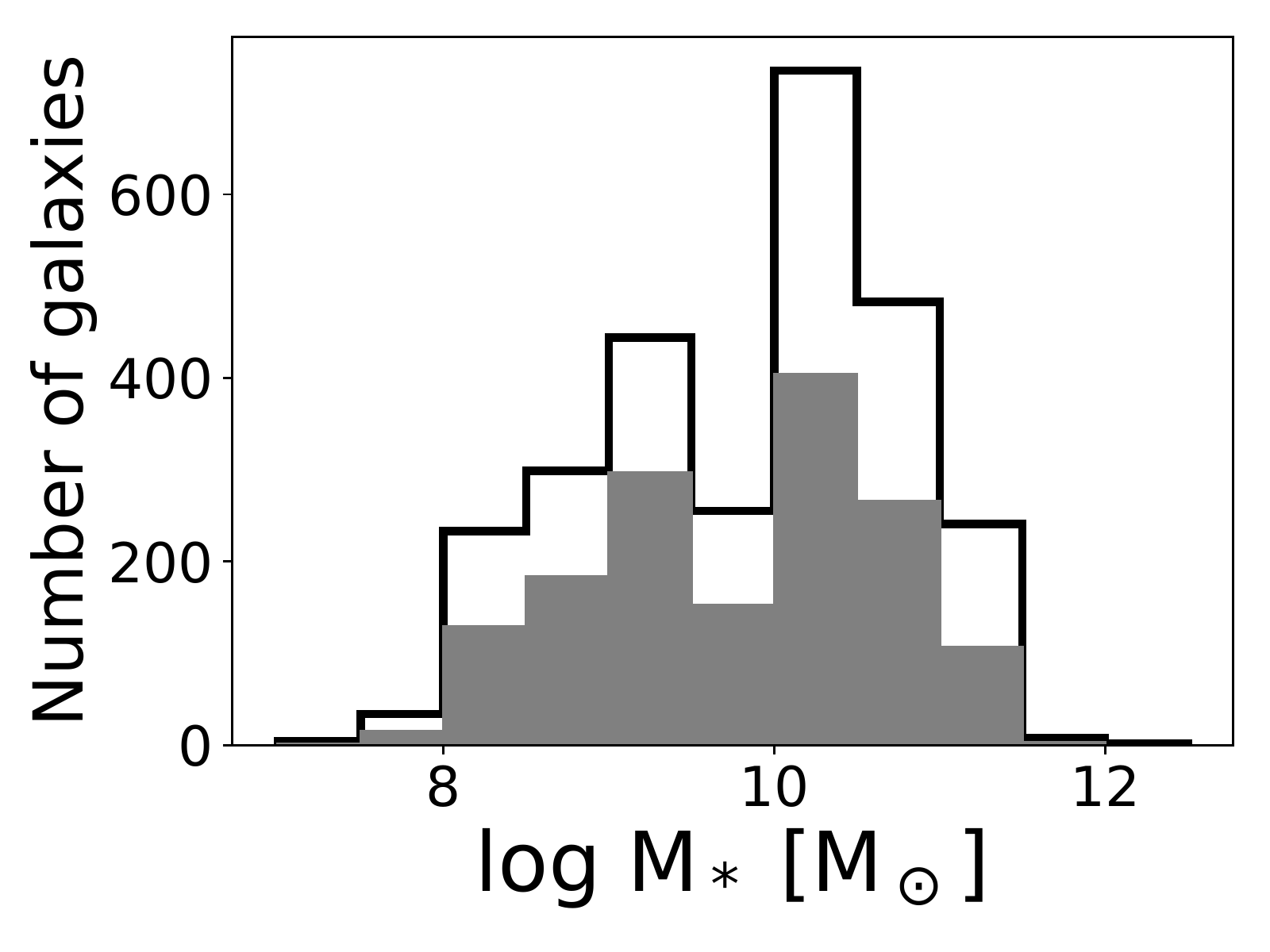}
\includegraphics[width=0.3\linewidth]{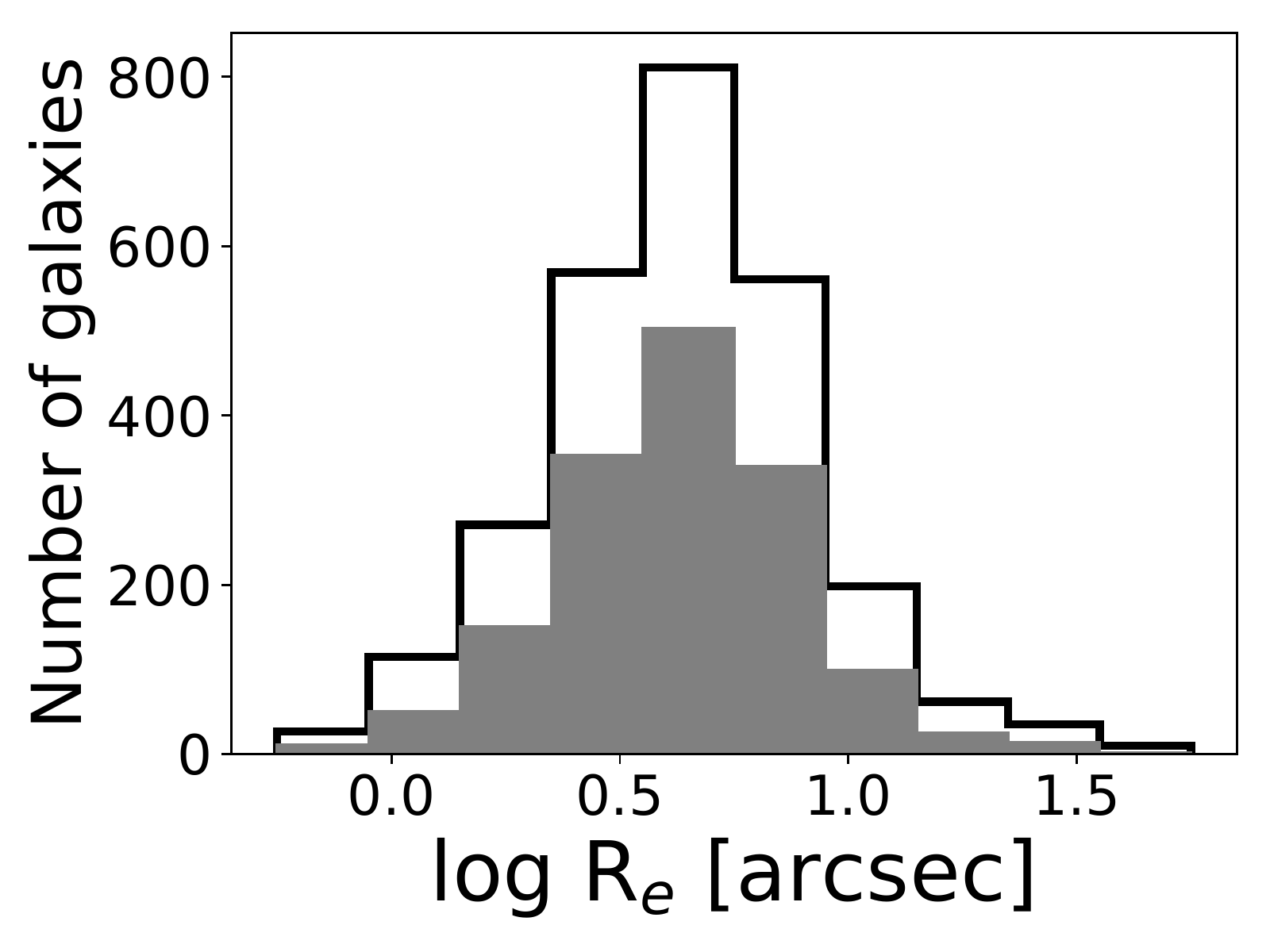}
\caption{Histograms showing the distribution of DR2 galaxy properties (gray histograms) with respect to the full SAMI-GAMA sample (clear histograms). From left to right the panels show the distribution with: redshift, log stellar mass \mstar\ and log effective radius \re. The DR2 sample is unbiased with respect to the complete, volume limited SAMI--GAMA parent sample.}
\label{fig:sample_properties}
\end{figure*}

Fibre flat-fielding in DR1 also used dome flat observations, which, as noted above, suffer from reduced counts at bluer wavelengths. In DR2 we instead used twilight sky observations (where available) that have significantly higher counts at the bluer wavelengths, compared to dome flats. In DR1, at wavelengths below $\sim 4000$\ \AA, variations in the fibre flat-field with fibre number were caused by unaccounted for scattered light (at the level of a few tens of counts). This issue resulted in large, unphysical variations in the fibre flat-field frame at blue wavelengths. The increased blue counts in the twilight sky frames largely eliminates this issue.  While the twilight sky has considerable spectral structure, once this is divided by the mean spectrum, the residual structure is small (a few per cent in the strongest spectral features, e.g.\ the Ca{\small II} H and K lines).  This remaining structure was removed by fitting a B-spline (with 16 knots positioned uniformly along the spectrum), including sigma clipping of outlying points. This effective smoothing of the fibre flat--field is appropriate as small-scale pixel-to-pixel differences in CCD response have already been removed at an earlier stage.  Therefore the fibre flat-field is only removing any residual difference in the slowly varying wavelength response of the system. Any individual outlying fibres were identified and replaced by comparison with a median stack of at least five twilight sky observations, before applying the fibre flat field to the science frames. Further, the colour response of the SAMI fibres is stable over an observing run, such that the RMS scatter between fibre flat-fields derived from twilight frames is 0.5 per cent or less (typically 0.2--0.3 per cent). 

The resampling of the data onto a calibrated wavelength axis has been modified in two ways that do not affect the quality of the wavelength calibration but instead improve the usability of the data. First, all SAMI Galaxy Survey data is now sampled onto a single, common wavelength scale, 3650 -- 5800\ \AA\ in the blue and 6240 -- 7460\ \AA\ in the red (with dispersions of 1.050\,\AA\,pixel$^{-1}$ and 0.596\,\AA\,pixel$^{-1}$ respectively). This uniformity facilitates the combining of data observed under different central wavelength settings without the need for a second resampling of the data. Secondly, at the time of resampling, the data are automatically corrected to a heliocentric frame. Both the heliocentric velocity correction and fixed wavelength range modifications are applied within the wavelength calibration step of data reduction, so no new interpolations of the data are required. The modified wavelength range compared to DR1 results in a very small (1--2\%) reduction in spectral sampling.

Finally, we note the recently discovered issue of weak charge--traps in the new (installed in mid-2014) red arm CCD of AAOmega \citep{aaoobschargetraps2018}.  There are a small number of partial rows (typically a few tens of pixels in length) that have shallow traps (typically a few tens of counts).  These are located near the top of the new AAOmega red CCD.  These have not been corrected in the current DR2 data, but will be in future releases.  The impact of these features in the current DR2 data is that for galaxies with data near the top of the detector (FITS header keyword IFUPROBE=1 in the final cubes), there can be a small number of spectra that show small (few tens of counts) dips in the final cubes.

\section{Data Release 2}
\label{sec:dr2_sample}
The SAMI Galaxy Survey Second Public Data Release (DR2) sample consists of 1559 unique galaxies. This sample represents all SAMI Galaxy Survey galaxies observed up to the 1$^\mathrm{st}$ July 2017 that lie in the GAMA regions of the survey and for which all value--added products have been derived. In addition we require that all galaxies satisfy a set of quality criteria.  Their data must consist of at least 6 observations (out of the 7 nominal dither positions) where each observation has i) a measured point spread function (PSF), derived from a Moffat profile fit, with Full Width at Half Maximum (FWHM) better than 3.1 arc seconds and ii) atmospheric transmission better than 55 per cent. These criteria result in 72 galaxies being excluded from DR2. One further observed galaxy is rejected due to being an ancillary target that was not part of the GAMA survey and so lacks important supporting photometric data. Of the 1632 galaxies eligible for DR2 we therefore reject 73, for a final sample of 1559 unique galaxies. This sample represents approximately a factor of 2 increase over DR1. The remaining galaxies will be made publicly available as part of a future data release.

The galaxies in DR2 span a broad range in stellar mass, \mstar, effective radius, \re, redshift and visual morphology. \mstar, \re\ and redshift (along with a number of other general galaxy properties) are provided by the GAMA survey \citep{driver2011,bryant2015}. Visual morphology classification has been performed taking advantage of SDSS DR9 {\it gri} colour images, as discussed in \citet{cortese2016}. Briefly, galaxies are first divided into late- and early-types according to the presence/absence of spiral arms and/or signs of star formation. Pure bulges are then classified as ellipticals (E) and early-types with disks as S0s. Similarly, late-types with only a disk component are Sc or later, while disk plus bulge late types are Sa--Sb. 

\begin{figure*}
\centering
\subfigure{\includegraphics[width=\columnwidth]{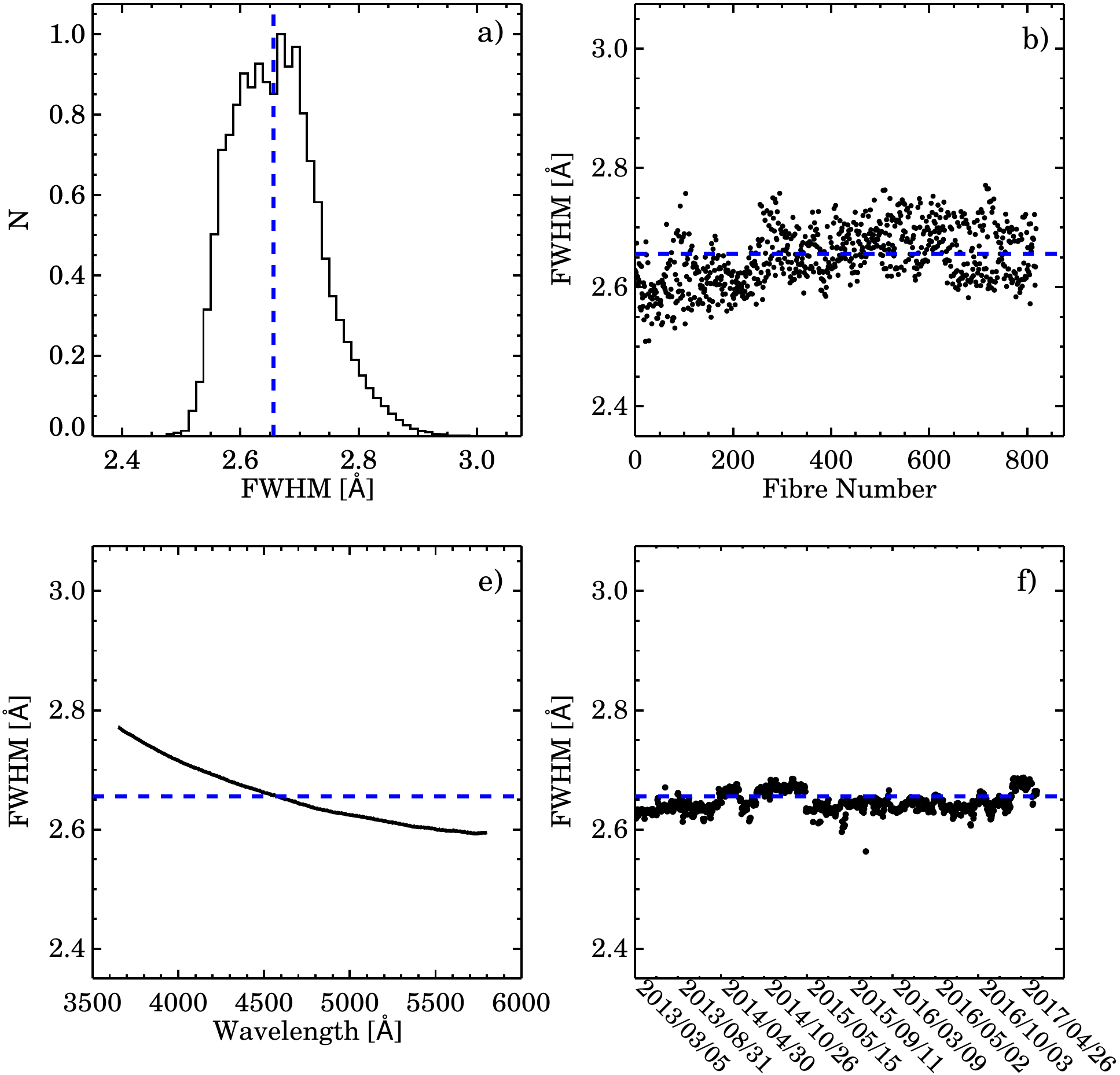}}
\subfigure{\includegraphics[width=\columnwidth]{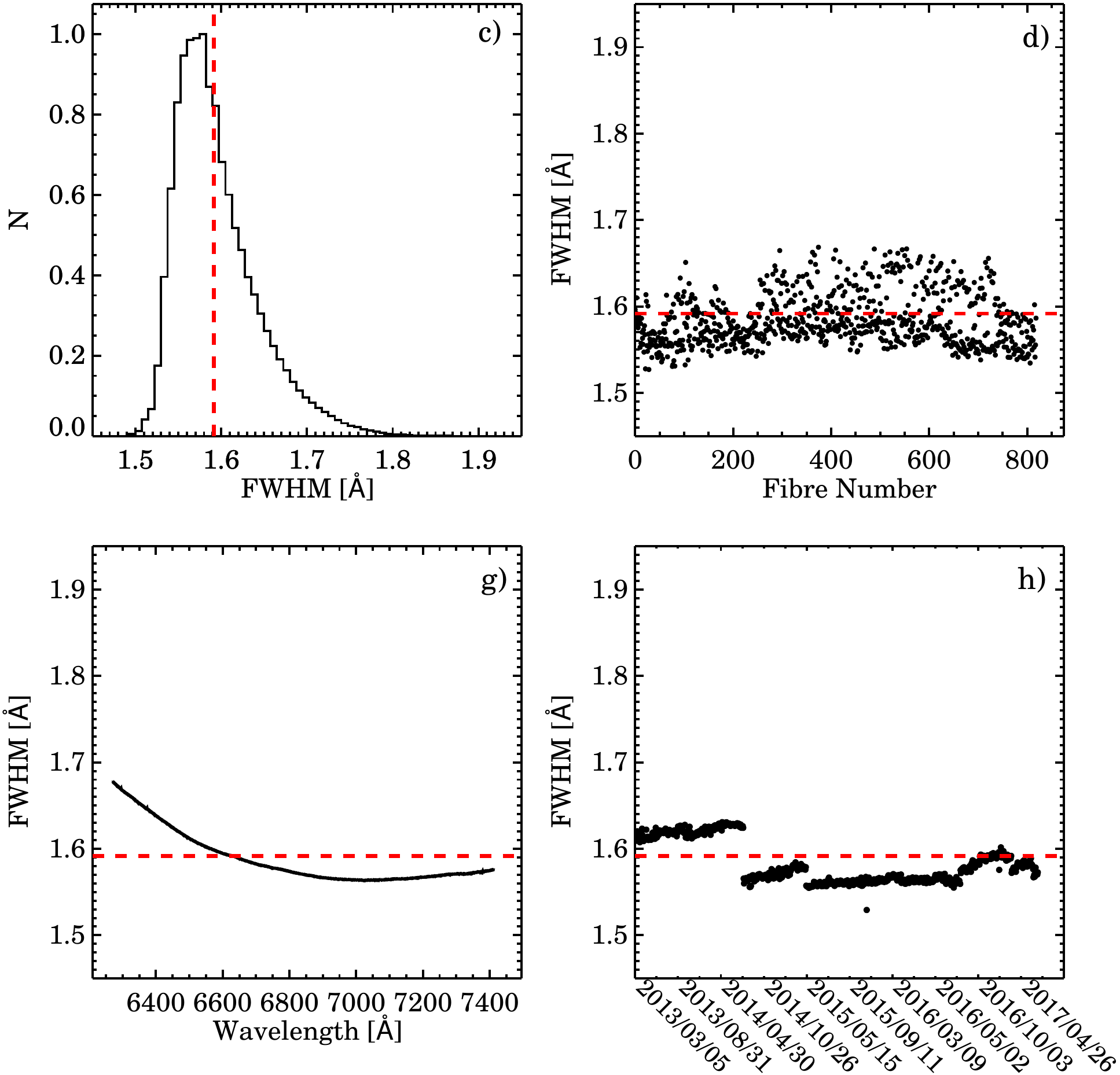}}
\caption{SAMI-AAOmega spectral resolution in the blue arm and the red arm derived from reduced arc-line frames. Panel a,b,e,f shows: the FWHM distribution, FWHM versus fibre number, FWHM versus wavelength, and FWHM versus date in the blue data (note that dates are linearly spaced on frame number, not time), and similar for Panel c,d,g,h for the red data.}
\label{fig:spectral_resolution}
\end{figure*}

\begin{table*}
	\centering
	\caption{SAMI spectral resolution parameters in blue and red. This table gives an overview of the wavelength range ($\lambda_{\rm{range}}$), central wavelength ($\lambda_{\rm{central}}$), median FWHM of the best-fitting Gaussian to the spectral instrumental LSF in \AA, the standard deviation of this Gaussian in \AA, the spectral resolution at $\lambda_{\rm{central}}$ (R$_{\lambda-\rm{central}}$), the velocity resolution (FWHM) in \kms\, ($\Delta v$), and the dispersion resolution ($1\sigma$) in \kms\, ($\Delta \sigma$).}
	\label{tab:specres_table}
	\begin{tabular}{c c c c c c c c } 
		\hline
        Arm & $\lambda_{\rm{range}}$ [\AA] & $\lambda_{\rm{central}}$ [\AA] & FWHM [\AA] & $\sigma$ [\AA] & R$_{\lambda-\rm{central}}$ & $\Delta v$ [\kms] &$\Delta \sigma$ [\kms] \\		
		\hline
        Blue & 3750-5750 & 4800 & $2.66_{-0.070}^{+0.076}$ & 1.13 & 1808 & 165.9 & 70.4 \\
        Red  & 6300-7400 & 6850 & $1.59_{-0.040}^{+0.049}$ & 0.68 & 4304 & 69.7  & 29.6 \\
		\hline
	\end{tabular}
\end{table*}


All votes (varying between 8 and 14 individuals) are then combined. For each galaxy, the morphological type with at least two thirds of the votes is chosen. If no agreement is found, adjacent votes are combined into intermediate classes (E/S0, S0/Sa, Sbc) and, if the two-thirds threshold is met, the galaxy is given the corresponding intermediate type. When no agreement is reached, a new round of classifications is performed. However, this time the choice is limited to the two types with the most votes during the first iteration, with the galaxy being marked as unclassified if agreement is still not reached. For galaxies in DR2, 1450 galaxies (93 per cent) have been successfully classified during the first step, 46 (3 per cent) required a second iteration and for 63 (4 per cent) no agreement was found even after the second iteration. 

Fig.~\ref{fig:sample_properties} shows that the DR2 sample is unbiased with respect to the SAMI--GAMA parent sample in stellar mass, effective radius and redshift. We do not show the comparison for morphology because morphological classifications are not available for the full parent sample. These general galaxy properties are provided in the DR2 sample table included in this release.

\subsection{Data quality}

\begin{figure*}
  \centering  
  \includegraphics*[trim={90 0 10 0},width=0.33\linewidth]{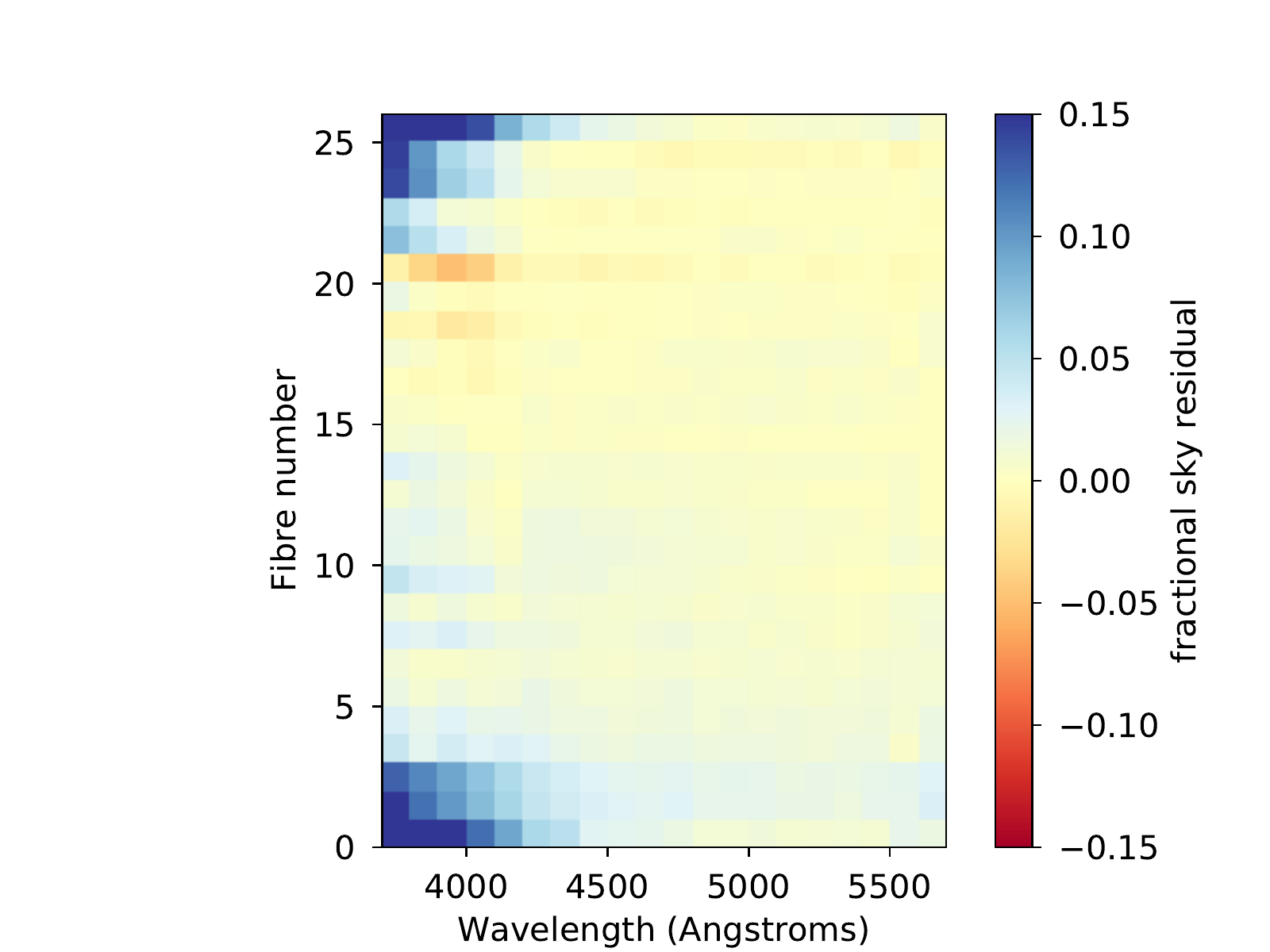}
  \includegraphics*[trim={90 0 10 0},width=0.33\linewidth]{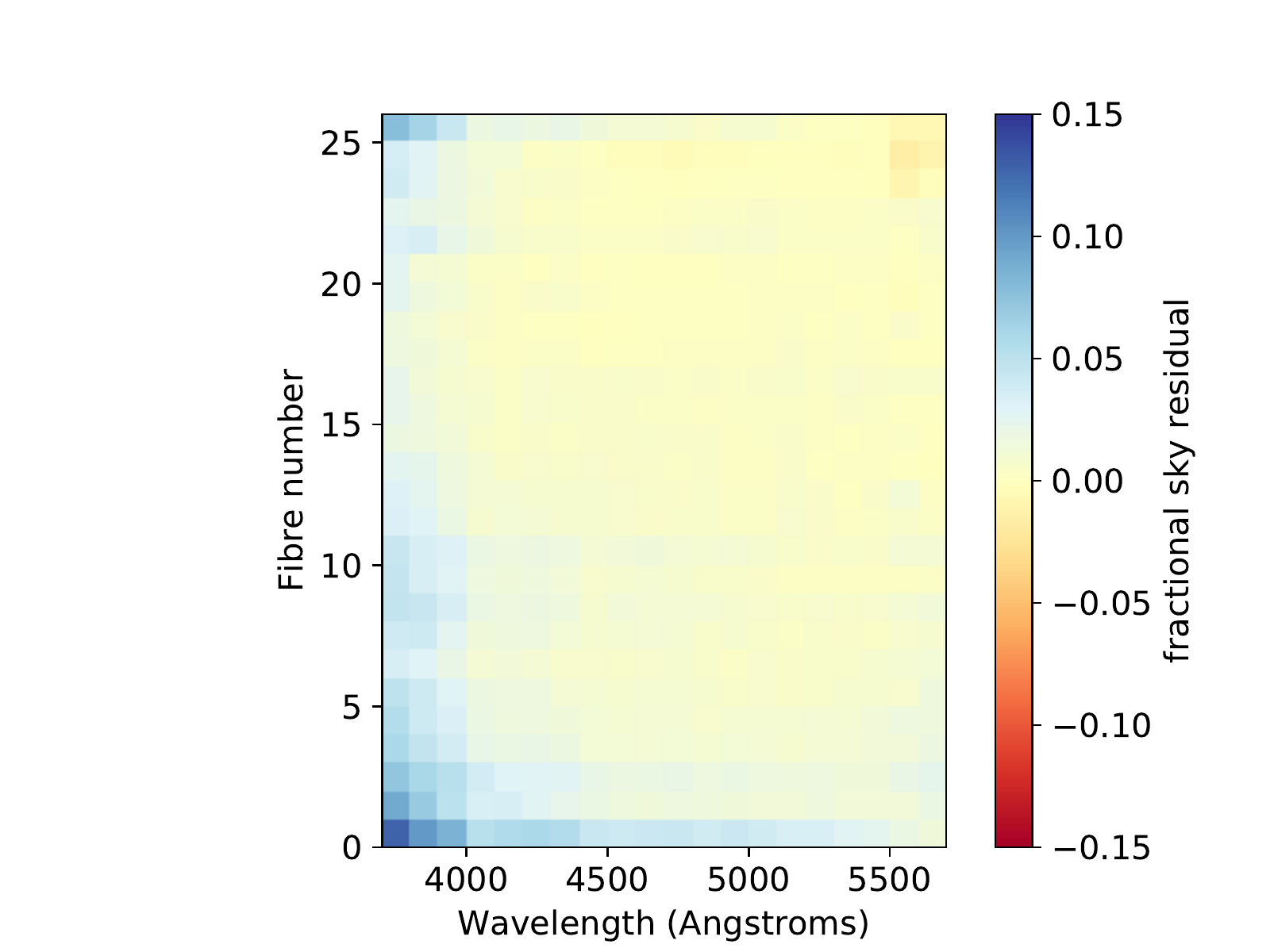}%
  \includegraphics*[trim={90 0 10 0},width=0.33\linewidth]{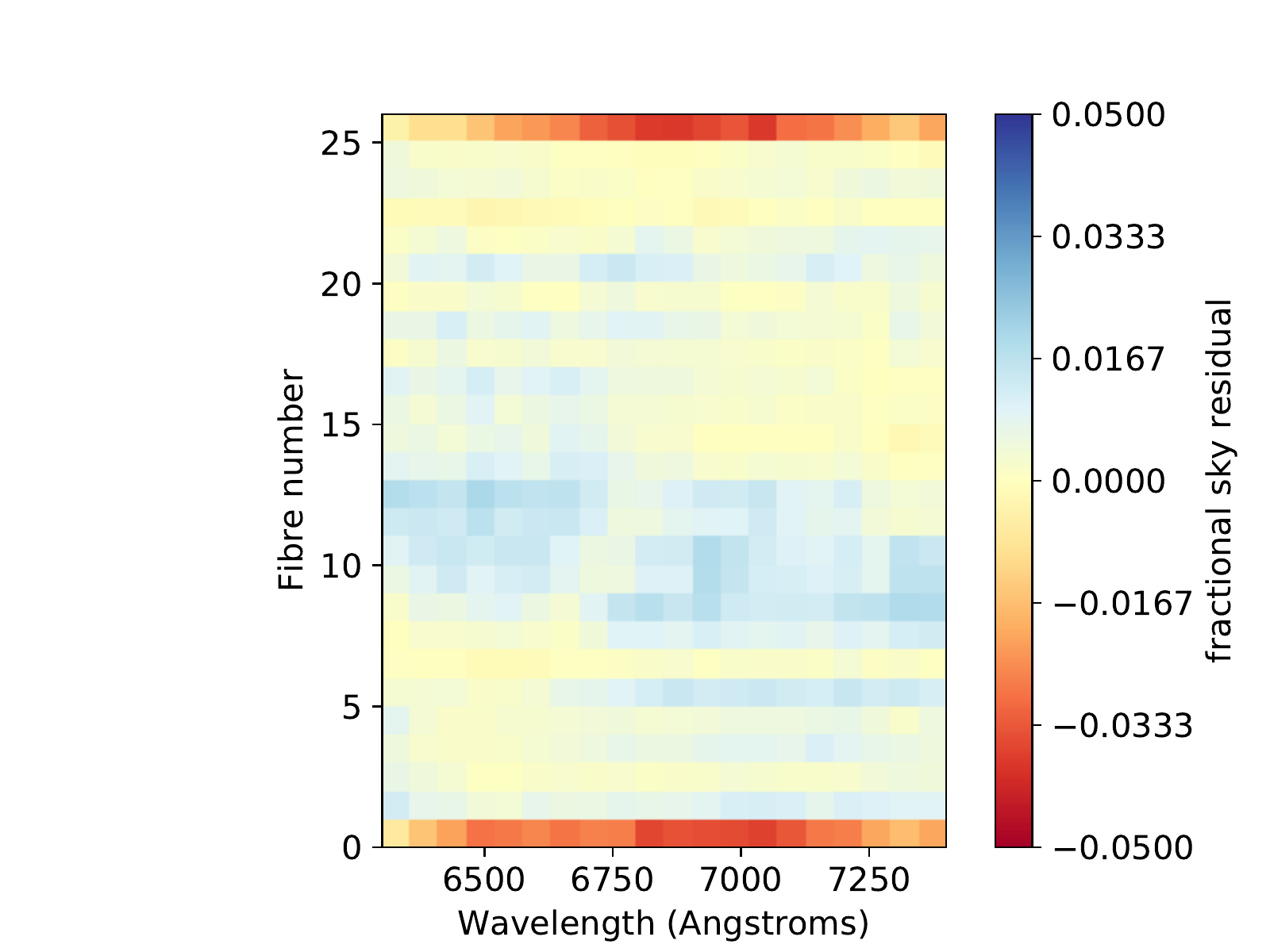}
  \caption{ %
    \label{fig:skysub}%
    The median fractional residuals in sky subtraction for SAMI sky fibres.  From left to right we show the old residuals from DR1 in the blue, the new residuals for DR2 in the blue, and the new residuals for DR2 in the red (unchanged from DR1).  For each sky
    fibre the flux is summed within 20 uniform bins in wavelength before determining the fractional residual.  The median residual within each bin is then calculated across all object frames within the survey.  Various low level systematic trends can be seen, including increased residuals for the sky fibres at the very edges of the slit.}
\end{figure*}

\subsubsection{Spectral resolution}
\label{subsec:spectral_resolution}

In this section we describe the spectral resolution as derived from SAMI-AAOmega data using reduced arc-line frames. We follow the method outlined in \cite{vandesande2017a} that has been implemented in {\sc 2dfDR} (Beta Version 6.65). The FWHM of the spectral instrumental line-spread-function (LSF) is derived using a Gaussian function, which is a good approximation for the SAMI-AAOmega LSF \citep{vandesande2017a}. {\sc 2dfDR} fits 24 unsaturated, unblended CuAr arc lines in the blue arm, and 12 lines in the red arm for all 819 fibres. The instrumental resolution over the entire wavelength region is derived from interpolating over individual arc-lines. Thus for every arc-line frame, we obtain a spectral resolution map of wavelength versus fibre number. We then calculate the spectral resolution maps for all 942 arc-line frames between 05/03/2013 to 26/09/2017. All data are combined into a three dimensional array with dimensions wavelength, fibre number, and observation date. In order to show the FWHM as a function of one dimension (e.g., wavelength, fibre number, or date), we will first collapse the three dimensional array along the two other dimensions using a median.

In Fig.~\ref{fig:spectral_resolution} we present the spectral resolution distributions, and the key resolution quantities for SAMI are given in Table~\ref{tab:specres_table}. We show the distribution of the spectral resolution in Fig.~\ref{fig:spectral_resolution}a,c, where we have taken the median along the fibre number dimension to reduce the number of FWHM values. We find that the distribution of the FWHM is more skewed in the red than the blue. There are small but significant resolution changes from fibre-to-fibre and with fibre position on the detector (Fig.~\ref{fig:spectral_resolution}b,d). In the blue, the resolution (FWHM) changes from $\sim 2.55$\ \AA, at the ``bottom" of the detector to $\sim 2.7$\ \AA\, two thirds up (fibre 600). In the red, the fibre-to-fibre resolution also changes with fibre position on the detector, from 1.55\ \AA\, for fibre 1 to a maximum of 1.65\ \AA\, around fibre $\sim500$.

We find a decrease in FWHM (increase in resolution) as a function of wavelength as shown in Fig.~\ref{fig:spectral_resolution}e,g. For the blue arm, the FWHM changes from 2.75\ \AA\ at 3700\ \AA\ to 2.6\ \AA\ at 5500\ \AA; for the red we find FWHM=1.65\ \AA\ at 6300\ \AA\ to 1.57\ \AA\ at 7000\ \AA, but then stays constant. Finally, in  Fig.~\ref{fig:spectral_resolution}f,h, we present the spectral resolution as a function of observing date. We find a change in the blue FWHM at the start of 2014, when the blue CCD was replaced (with an identical CCD, but with fewer cosmetic defects).  However, the change is small ($\sim1$ percent) and no greater than other changes at other times. In the red arm, we see a drop of $\sim4$ percent in the FWHM starting from October 2014 onwards. This drop coincides with the time when the red CCD was upgraded.

In summary, in the blue arm, we find a median resolution of: FWHM$_{\rm{blue}}= 2.66$\ \AA, and in the red arm of: FWHM$_{\rm{red}} = 1.59$ \AA. The fibre-to-fibre FWHM variation is 0.048\ \AA\ (RMS scatter) in the blue and 0.030\ \AA\ in the red. Over a period of four years, we find FWHM variations of 0.016\ \AA\ in the blue arm, and 0.024\ \AA\, in the red arm. The FWHM decreases with increasing wavelength in the blue arm by 0.051\ \AA, and red arm by 0.031\ \AA. 

\begin{figure*}
\centering
\includegraphics[clip,trim=0 0 0 0,width=0.5\linewidth]{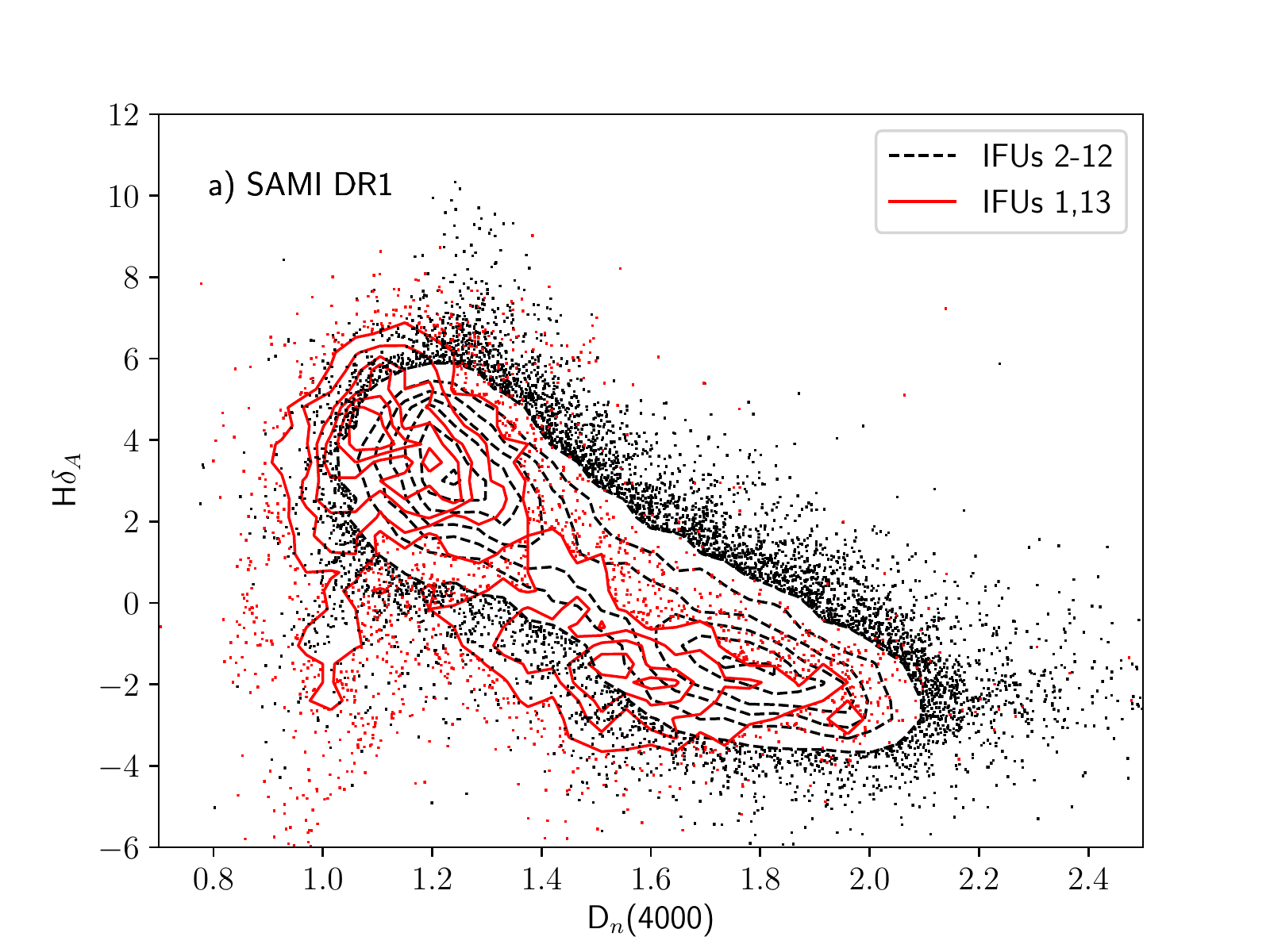}\includegraphics[clip,trim=0 0 0 0,width=0.5\linewidth]{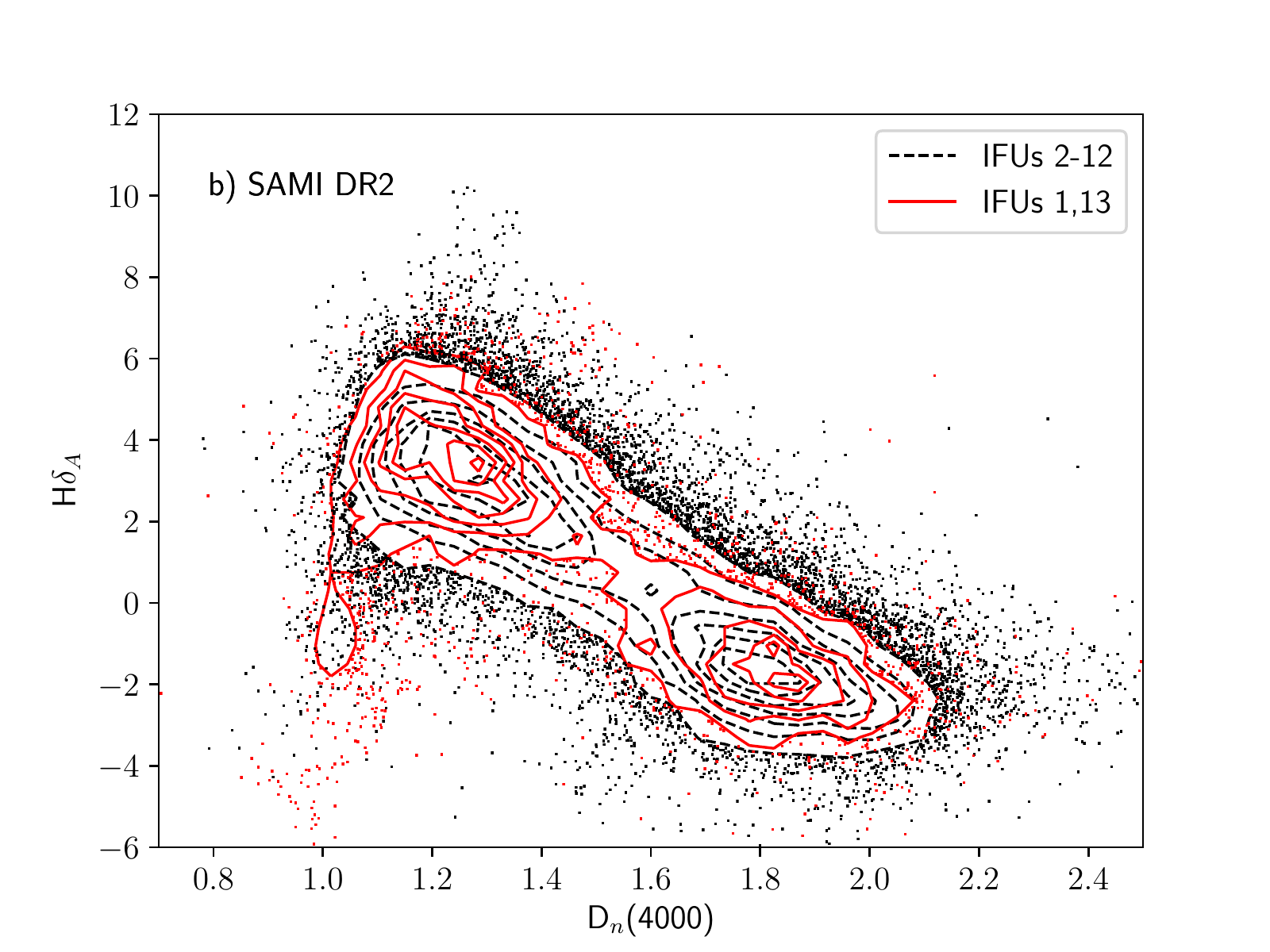}
\caption{The distribution of SAMI spaxels with median blue arm $S/N>10$ in the D4000$_n$ vs.\ H$\delta_A$ plane for a) data in SAMI DR1, and b) data in SAMI DR2.  Only spaxels from galaxies that are common to both DR1 and DR2 are shown.  IFUs 2--12 (black dashed contours and points) and IFUs 1, 13 (red contours and points) are shown separately.  The improved DR2 data reduction leads to consistent locations in the D4000$_n$ vs.\ H$\delta_A$ for all IFUs.  Contours are linearly spaced in point density.}
\label{fig:d4000_hdelta}
\end{figure*}

\begin{figure}
\centering  
\includegraphics[width=1.0\linewidth]{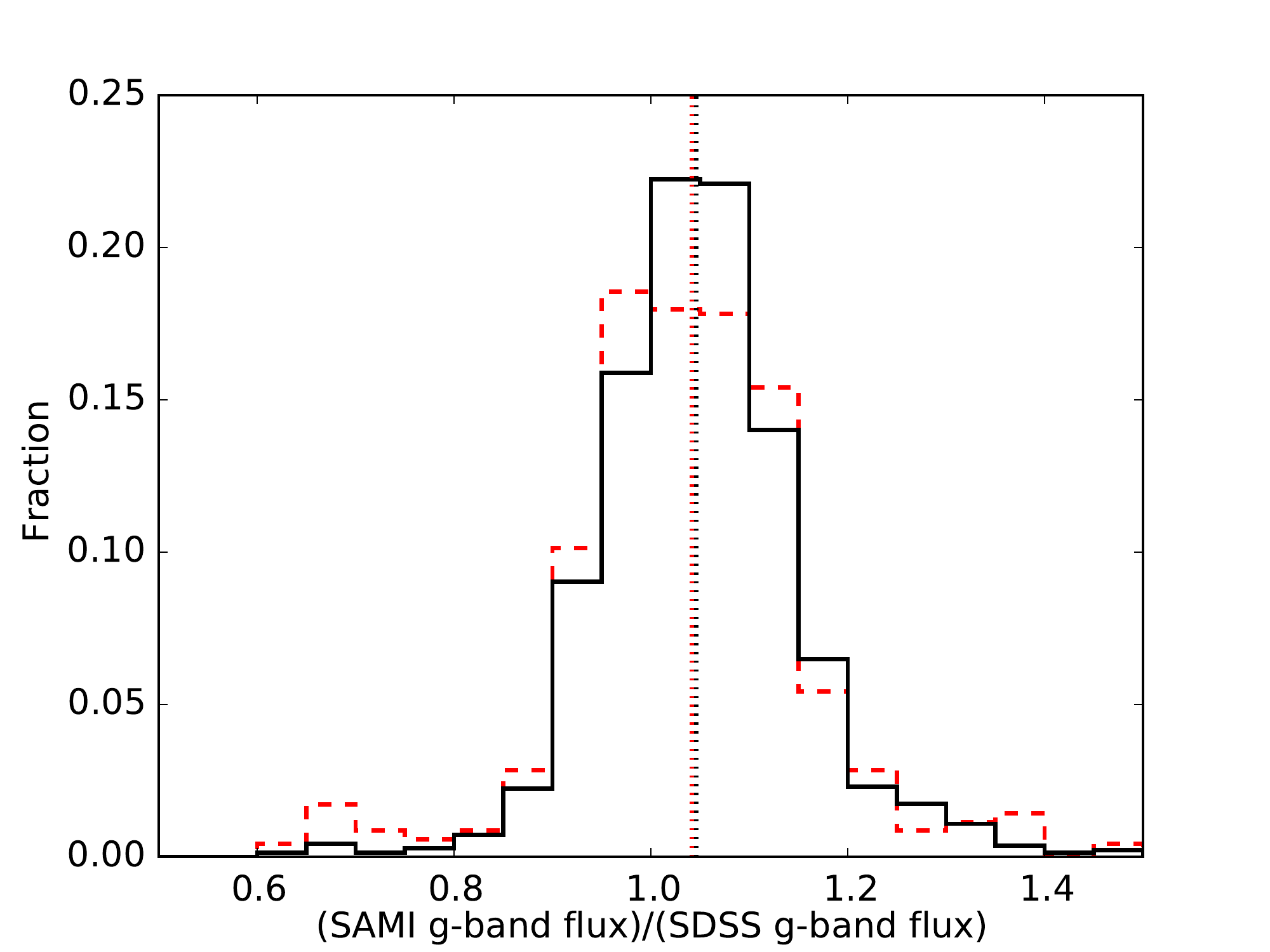}
\caption{The distribution of flux ratios between SAMI cubes and SDSS images in the $g$--band using 8 arcsec diameter circular apertures.  We compare SAMI DR1 (red dashed lines) and DR2 (solid black lines), with the vertical dotted lines showing the median flux ratio for each sample. Each histogram is normalized to the number of objects in the sample.}
\label{fig:fluxcal}
\end{figure}

\begin{figure*}
\centering
\includegraphics[clip,trim=10 10 10 10,width=0.3\linewidth]{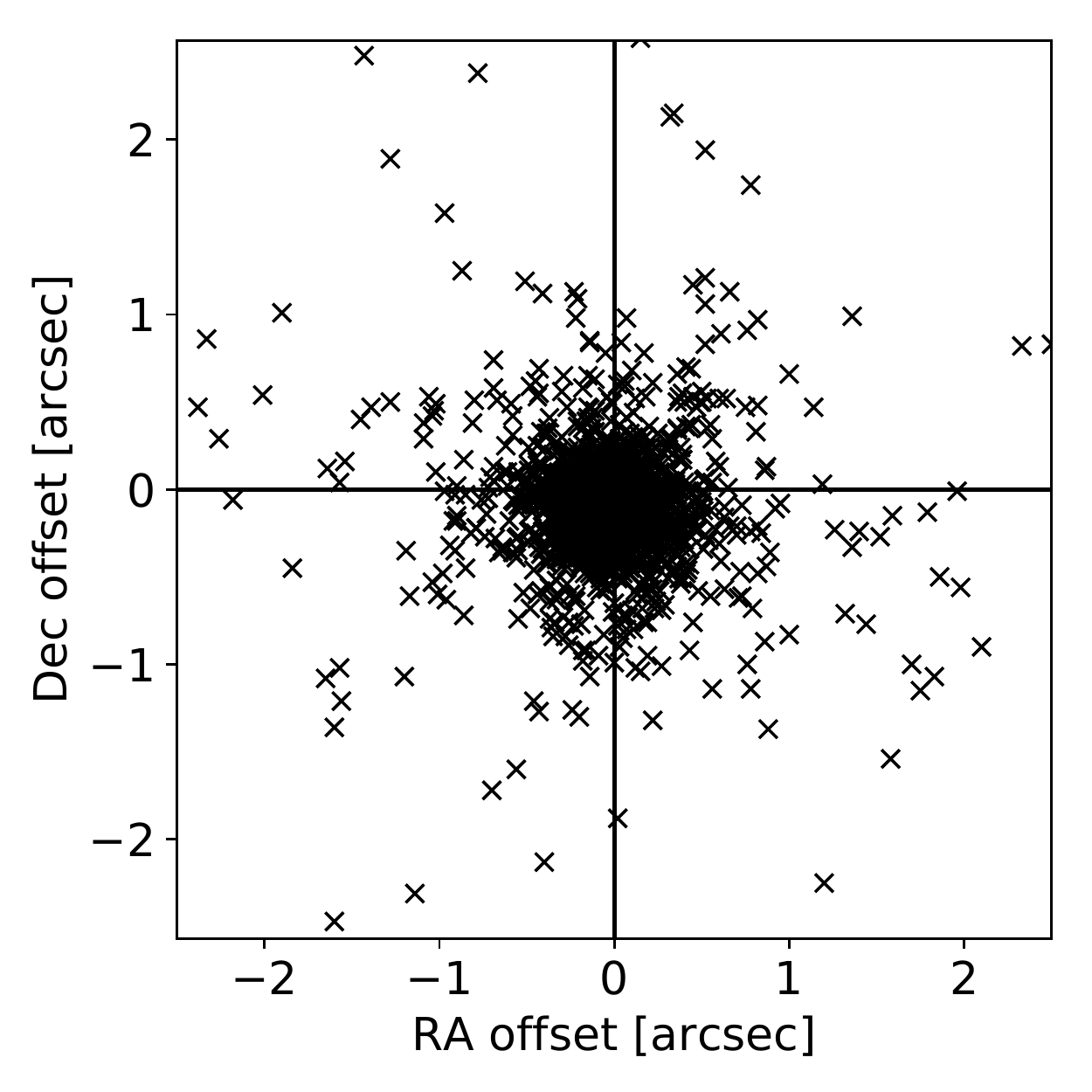}
\includegraphics[clip,trim=10 -30 10 10,width=0.33\linewidth]{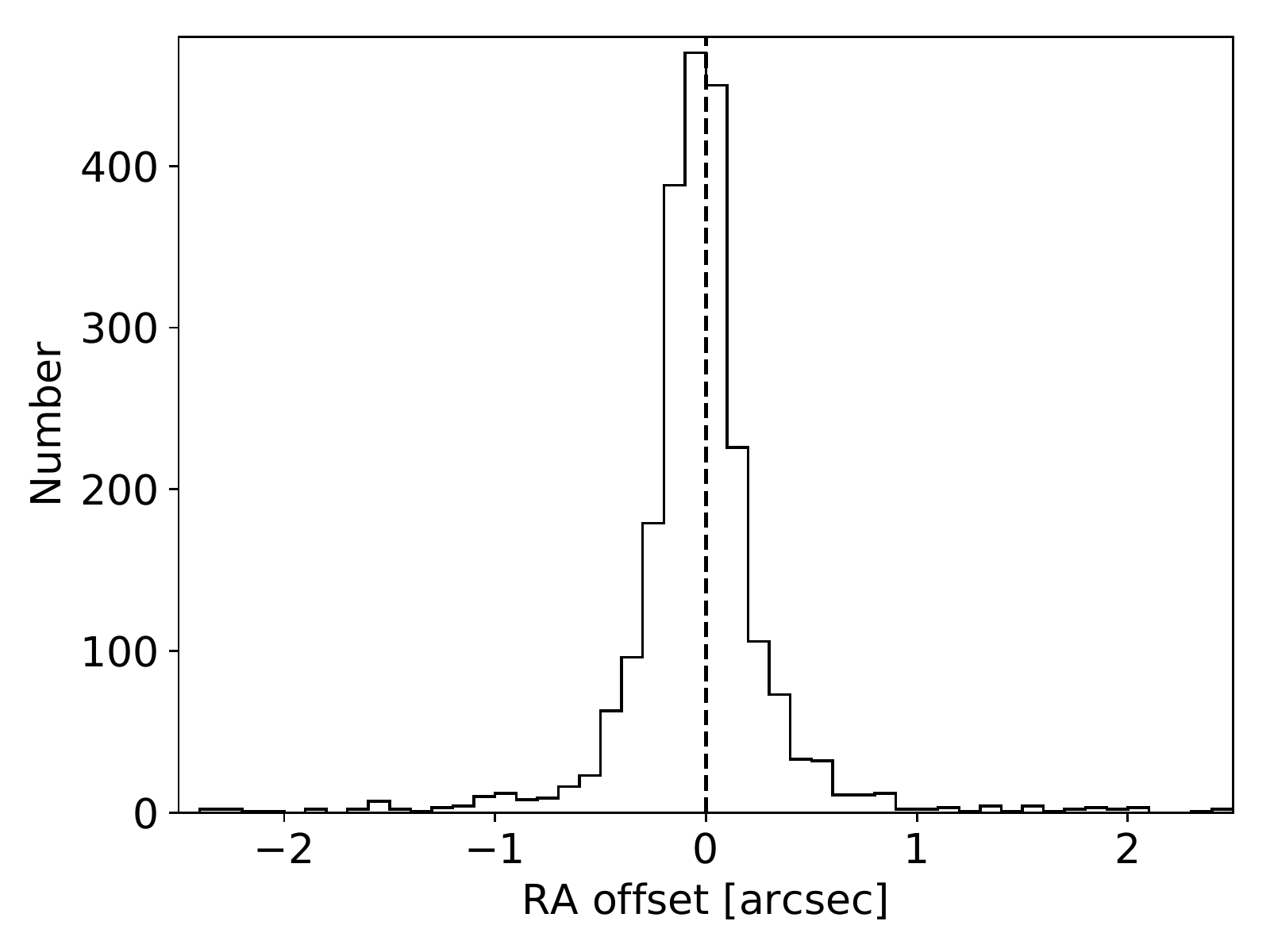}
\includegraphics[clip,trim=10 -30 10 10,width=0.33\linewidth]{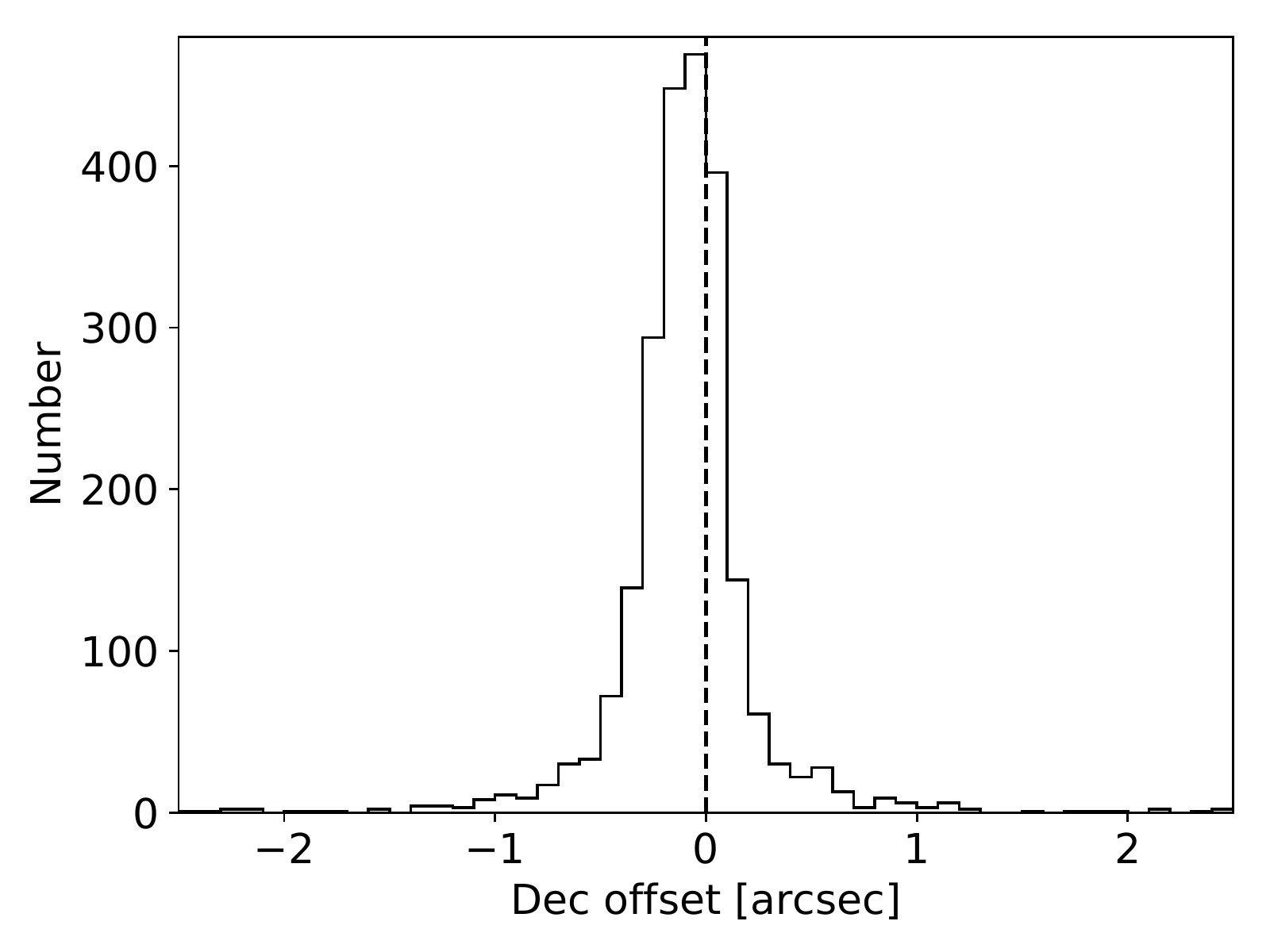}
\caption{Offset between the SAMI cube WCS and KiDS $r$--band imaging. The left panel shows the offsets for all DR2 galaxies. Note that 24 galaxies with large offsets lie outside the figure. The centre and right panel show histograms of the offset in RA and Dec respectively. Catastrophic failures have not been removed.}
\label{fig:wcs}
\end{figure*}

\subsubsection{Sky subtraction accuracy}
\label{subsec:sky_subtraction_accuracy}

The improvements to profile measurement and fibre flat fielding outlined in Section \ref{sec:dr_changes} lead to a substantial reduction in systematic sky subtraction residuals, particularly in the blue arm of the spectrograph. In Fig.\ \ref{fig:skysub} we show the median fractional sky subtraction residuals across 1750 individual data frames (for exposures of at least 900s), in 20 wavelength bins for each sky fibre in each arm of the spectrograph. The sky fibres are uniformly distributed along the slit, so systematic variations with sky fibre number relate to variations along the slit. There is no change in the red arm residuals between DR1 and DR2. The DR2 residuals in the blue arm (Fig.\ \ref{fig:skysub}, centre) are significantly reduced compared to the same measurement from DR1 (Fig.\ \ref{fig:skysub}, left), particularly at the corners of the CCD. This reduction is because the new approach to fibre flat--fielding reduces the impact of ghost features present at these locations in the AAOmega spectrograph.  Previously these features were at the level of up to $\sim20$\%, but are now reduced to $\sim5$\% or less. Other weak systematic features remain, including a gradient at the level of $\sim1$\% from top to bottom of the CCD, higher residuals for the slit end fibres (fibres 1 and 26) and an increase for all fibres at the blue end. This last feature is largely driven by reduced S/N at the blue end of the blue arm, rather than any actual systematic reduction in sky subtraction accuracy.

We perform a second test of DR2 data quality that examines our improved spectral extraction and sky subtraction (see Section \ref{sec:dr_changes}).  As we saw in Fig. \ref{fig:skysub}, there is a significant improvement in sky subtraction below 4000\AA\ at the top and bottom of the detector.  This most strongly influences IFUs number 1 and 13 that are located at either end of the spectrograph slit.  To demonstrate the improvement we measure the D4000$_n$ \citep{1999ApJ...527...54B} and H$\delta_A$ \citep{1997ApJS..111..377W} indices for all spaxels with median $S/N>10$ from galaxies that are contained within both DR1 and DR2.  For SAMI DR1 the data in IFUs 1 and 13 (at the ends of the slit) show a systematic offset to lower values of D4000$_n$ (red contours and points in Fig.\ \ref{fig:d4000_hdelta}a).  In contrast, the same spaxels in SAMI DR2 are completely consistent with the distribution in the other IFUs (Fig.\ \ref{fig:d4000_hdelta}b).  We note that in this test we have not corrected the indices for emission lines, so a small number of spaxels lie at lower H$\delta_A$ values than might otherwise be expected (e.g.\ red points at D4000$_n\simeq1$ and H$\delta_A\simeq-4$). 

\subsubsection{Flux calibration}
\label{sec:fluxcal}

We compare the flux calibration of SAMI DR2 data to SDSS $g$--band images.  The same procedure as described for the DR1 sample in \citet{green2018} is also carried out on DR2.  This procedure compares fluxes in SDSS $g$--band images and SAMI cubes within an 8 arcsec diameter aperture. The SAMI cubes are convolved with the SDSS $g$--band filter curve and the SDSS images are convolved to the median seeing of SAMI.  Galaxies with integrated fluxes below 100 $\mu$Jy were not included, to avoid extra scatter from low S/N. The median flux ratio (SAMI/SDSS) is $1.048\pm 0.003$ (where the error is the uncertainty on the median, not the RMS scatter), consistent with results from DR1. As can be seen in Fig. \ref{fig:fluxcal}, the distribution of flux ratios is slightly narrower for DR2 (solid line) than DR1 (dotted line). 95 per cent of objects have a flux ratio within $\pm 0.15$ of the median. Regarding the accuracy of relative flux calibration, this is unchanged compared to previous data releases.  As noted in \citet{allen2015}, we find a colour offset, $\Delta(g - r)$, of 0.043 with a standard deviation of 0.040, with respect to the SDSS PSF magnitude derived colours of the SAMI secondary standard stars.

\subsubsection{WCS accuracy}
\label{sec:wcs}

During cube construction we register the galaxy centroid in each individual dither by fitting a two dimensional Gaussian to the observed flux. The dithers are aligned using these centroids, and are then combined such that the galaxy centre is located at cube spaxel coordinates (25.5, 25.5). We then assign the catalogue right ascension (RA) and declination (Dec) of the galaxy to this spaxel coordinate and define the World Coordinate System (WCS) of the cube relative to this position. While accurate for the majority of galaxies, there remains some uncertainty in the WCS due to the centroiding process, and, in a limited number of cases, the pipeline can misidentify the galaxy centre resulting in a significant offset.

We verify the accuracy of the WCS of the data by visual inspection and matching to $r$--band images from the Kilo Degree Survey \citep[KiDS;][]{dejong2017}. In Fig. \ref{fig:wcs} we show the offset in RA and Dec between the centre of the collapsed SAMI cube and the centre of the galaxy in the KiDS image. We find the mean offset is $-0.016 \pm 0.020$ arcsec in RA and $-0.102 \pm 0.017$ arcsec in Dec. When we remove catastrophic failures (see below), the mean offset changes to $-0.027 \pm 0.009$ in RA and $-0.106 \pm 0.008$ in Dec, resulting in a decreased rms scatter, however the small statistically significant offset remains.

For cubes where multiple galaxies or a foreground star is present in the hexabundle  or the galaxy is highly structured, the simple two-dimensional Gaussian fit can misidentify the galaxy centre, resulting in a large positional offset between the cube centre and the true galaxy centre. After visual inspection of the cubes, we determined that 50 galaxies suffered from significant offsets ($> 1$ arcsec in radial offset) due to these issues. For these galaxies, we shift the cube WCS to match that determined from the KiDS imaging. 

\subsubsection{Seeing distribution}

\begin{figure}
\centering
\includegraphics[width=\linewidth]{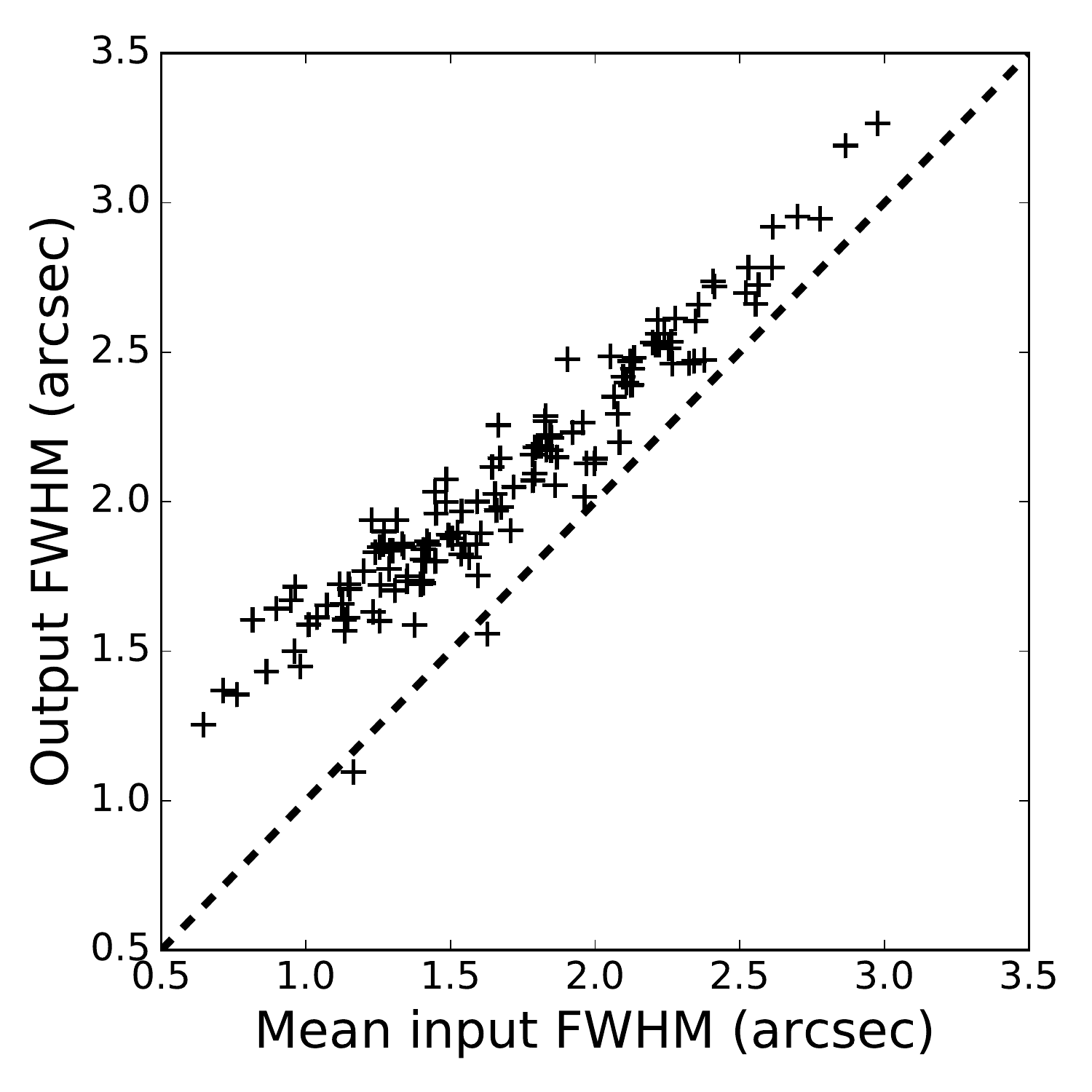}
\includegraphics[width=\linewidth]{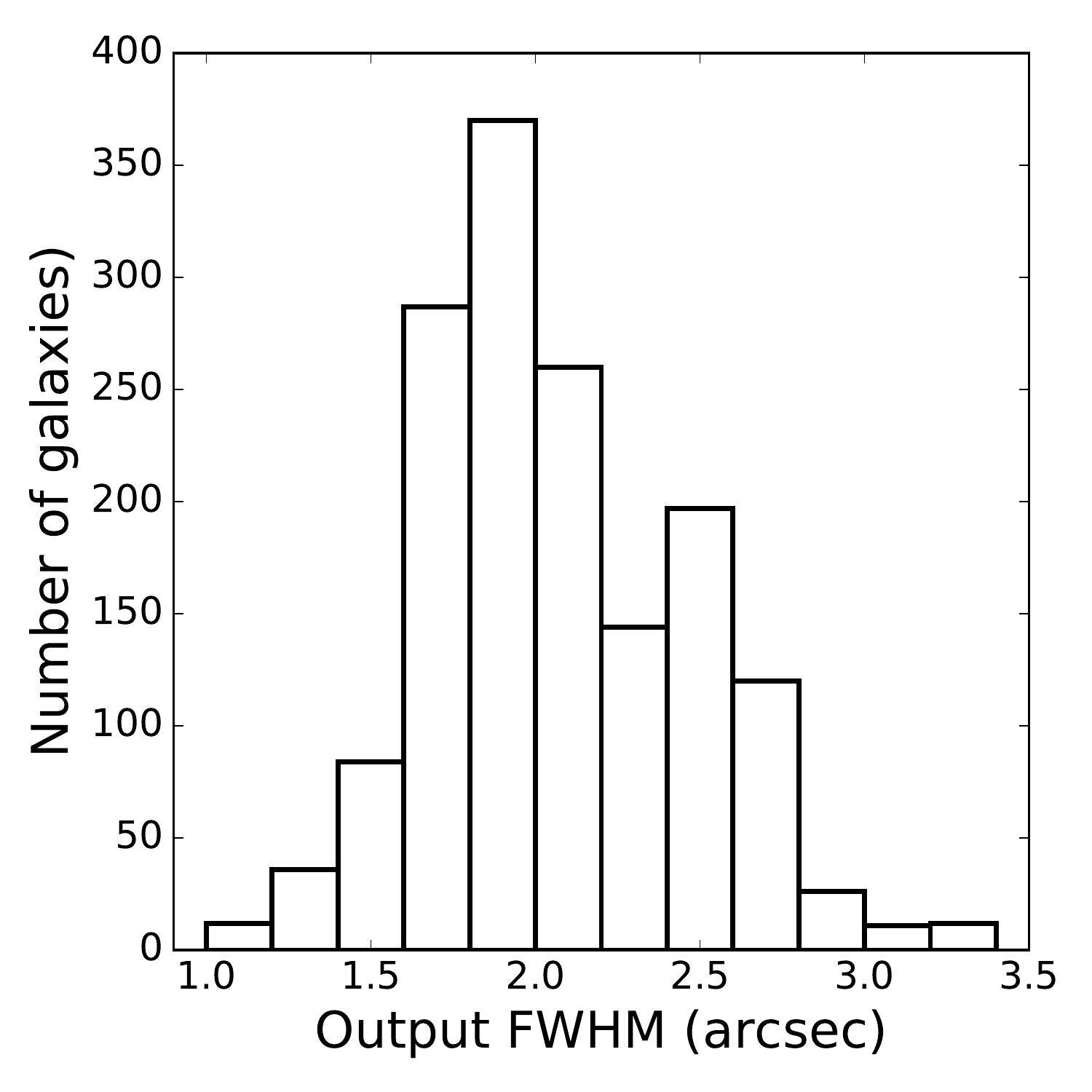}
\caption{Upper panel: comparison of the mean measured FWHM of individual dithered exposures to the FWHM of the reconstructed secondary standard star cubes. FWHMs are from Moffat-profile fits. The dashed black line is the 1:1 relation. Lower panel: distribution of the FWHM in the reconstructed cube for all SAMI DR2 galaxies. The mean of the FWHM is 2.06 arcseconds, and the standard deviation is 0.40 arcseconds.}
\label{fig:fwhm}
\end{figure}

Each observation has an associated PSF, characterised by the FWHM. For individual observations these are measured by fitting a Moffat profile to the flux distribution of the secondary standard star. The output data cubes also have a PSF that depends on the PSFs of the input observations, as well as the accuracy of registering these inputs to a common coordinate system. In the upper panel of Fig. \ref{fig:fwhm} we show the FWHM of the PSF of the cube of the secondary standard star as a function of the mean FWHM of the input observations. For mean input FWHM $\gtrsim 1.5$ arcsec, the input and output FWHM are linearly related, with the PSF of the cube being $\sim 0.2$ arc seconds broader than the mean input PSF. In good seeing, the difference between the cube PSF and the input PSF  is increased.  This is caused by two effects; the additional broadening in the output PSF due to uncertainties in the centroids of each input frame and the effect of the optical fibres, whose finite size effectively imposes a minimum FWHM on the PSF, even if atmospheric seeing is ignored. Atmospheric broadening is still the most significant factor in determining the FWHM of the output PSF.

In the lower panel of Fig. \ref{fig:fwhm} we show the distribution of FWHM for all galaxies in DR2, determined from Moffat profile fits to the secondary standard star cubes observed simultaneously with the galaxies. The mean FWHM of the output cubes is 2.06 arcseconds, varying between 1.10 and 3.27 arcseconds. 84 per cent of galaxies have FWHM better than 2.5 arcseconds.

\section{Core data}
\label{sec:core_data}
The data in SAMI DR2 is broadly divided into core data, produced directly from the SAMI data reduction pipeline, and value-added data products, derived from SAMI Galaxy Survey science analysis pipelines. Here we describe the core data, with the data products being described in Sections \ref{sec:emission_line_products} and \ref{sec:absorption_line_products}.

\subsection{Cubes}
\label{subsec:cubes}
The primary data produced by the SAMI Galaxy Survey are pairs of spectral data cubes for each observed galaxy, covering the blue and red part of the optical wavelength range. Each data cube consists of 2048 spectral slices, where each slice is a $50 \times 50$ square area of spatial pixels (spaxels). The sampling of the spatial axes is 0.5 arc seconds. For the blue cube the spectral sampling is 1.050\ \AA/pixel with a spectral FWHM of 2.66\ \AA, covering the wavelength range 3650 to 5800\ \AA. For the red cube the spectral sampling is 0.596\ \AA/pixel with a spectral FWHM of 1.59\ \AA, covering the wavelength range 6240 to 7460\ \AA. See Table \ref{tab:specres_table} for further details. In addition to the measured fluxes, each cube contains the variance, weight map and compressed covariance -- see \citet{sharp2015} for details. 

\subsection{Binned cubes}
\label{subsec:binned_cubes}
To complement the default cubes, we provide a set of three pre-binned data cubes, that we refer to as `adaptive', `annular' and `sectors'.  
\begin{itemize}
\item Adaptive: Bins are adaptively generated to contain a target S/N of 10, using the Voronoi binning code of \citet{cappellari2003}. The S/N is calculated from the flux and variance spectra of each spaxel as the median across the entire blue wavelength range. Spaxels with S/N $> 10$ are not binned.
\item Annular: Bins are generated as a series of elliptical annuli, centred on the centre of the cube. The position angle, PA, and ellipticity, $\epsilon$, of the galaxy are determined using the {\it find\_galaxy} {\sc Python} routine of \citet{cappellari2002} from the image generated by summing the cube along its wavelength axis. The spaxels are then allocated to five linearly-spaced elliptical annuli, each with the PA and $\epsilon$ of the whole galaxy.
\item Sectors: Bins are generated as a series of elliptical annuli, with each annulus further subdivided azimuthally into 8 regions of equal area. The axes of the sectors are defined in reference to the PA of the galaxy. The annuli are generated as for the annular binning scheme.
\end{itemize}
For each binned cube, we first generate a bin mask from the blue cube using the criteria described above, then sum the spectra for each spaxel contributing to a given bin to generate the binned spectrum. The bin masks generated from the blue cubes are applied to the red cubes as well to allow direct comparison. The output binned data cubes consist of $50 \times 50 \times 2048$ arrays, maintaining consistency with the original cubes. Each spaxel in the output cube contains the binned spectrum for the bin that it belongs to -- spaxels from the same bin contain identical spectra. All spaxels containing flux are allocated to a bin. This procedure is repeated for the variance cube, accounting for the covariance between spaxels in each bin. We note that the variance of large bins ($\gtrsim 25$ spaxels) may be underestimated by up to 5\% due to a small component of unaccounted-for covariance between included spaxels (this will be corrected in future releases). Each binned cube consists of the flux and variance cubes and an additional bin mask image, indicating which spaxels have been combined into each bin. All binned data are generated using the {\it binning} module of the SAMI data reduction pipeline. Variations on the adaptive and annular/sectors binning schemes can easily be generated by modifying functions within this module.

\subsection{Aperture spectra}
\label{subsec:aperture_spectra}

To facilitate comparison to existing single aperture surveys, we also provide a set of aperture spectra derived from the SAMI cubes. These aperture spectra are generated using the {\it binning} module of the SAMI data reduction pipeline as single-bin binned spectra, with two exceptions: spaxels lying outside the aperture are not allocated to a bin, and the flux of the aperture spectrum is re-scaled to account for the difference in area between the included spaxels and the true bin area. The data format is also different; for each aperture we generate a one dimensional flux array, a one dimensional variance array (accounting for spatial covariance between contributing spaxels), and a two dimensional bin mask image, indicating which spaxels have been summed to form the aperture. As for the binned data cubes, large apertures may have their variance underestimated by up to $\sim 5$ \%.

We provide six apertures, four of which are circular apertures centred on the centre of the cube with diameter 1\farcs4, 2\arcsec, 3\arcsec\ and 4\arcsec\ respectively. We provide a fifth circular aperture with fixed physical diameter of 3kpc, determined using the observed redshift of the galaxy from the GAMA survey and our adopted cosmology. The sixth aperture is an elliptical aperture of major axis radius \re, where $\epsilon$, PA and \re\ are taken from v09 of the GAMA S\'{e}rsic catalogue \citep{kelvin2012}.

\subsubsection{S/N}
\label{sec:ap_sn}
\begin{figure}
\includegraphics[width=\linewidth]{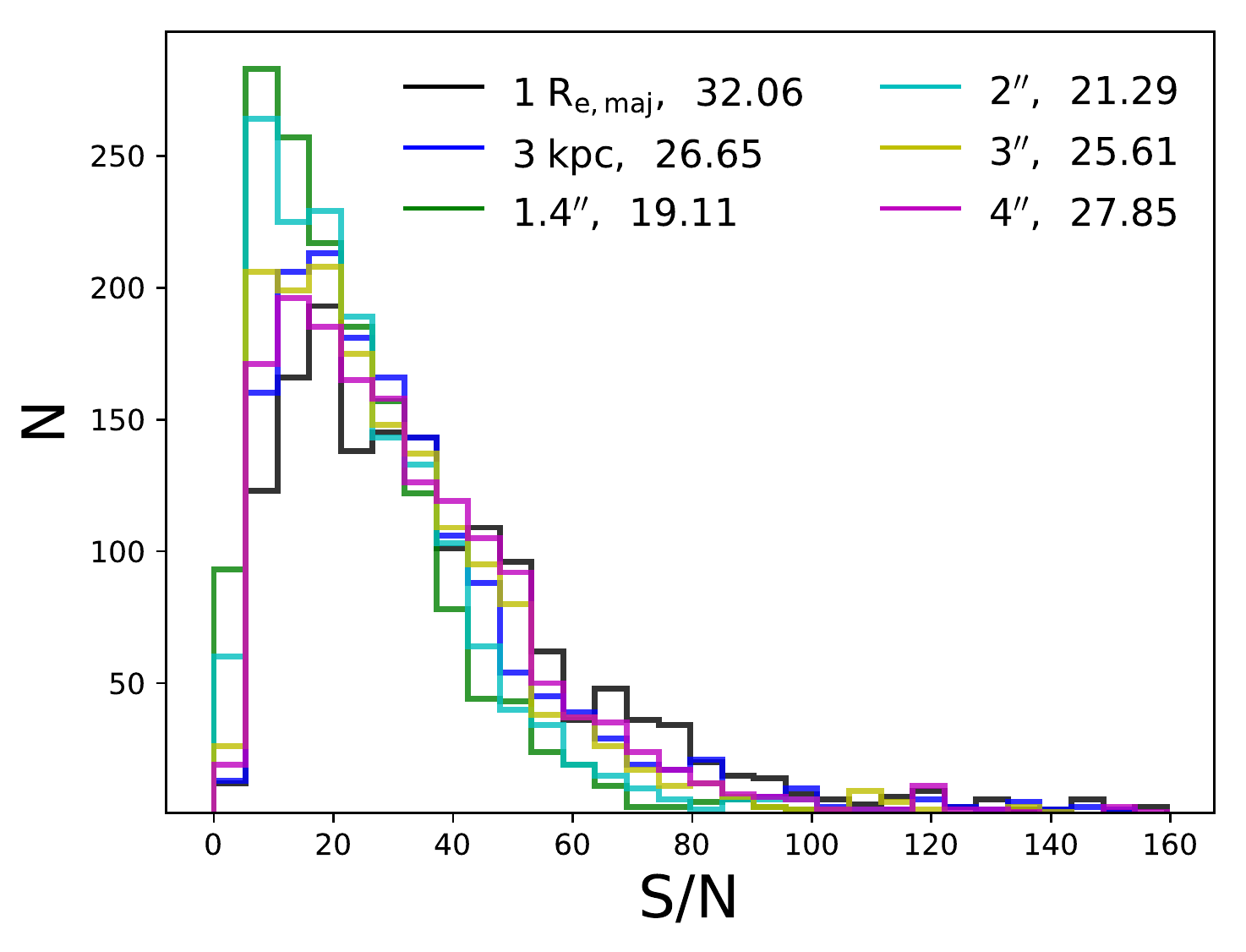}
\caption{Histograms of the aperture S/N for all DR2 galaxies, in the six available apertures. The S/N is the median value in the range from 4600 to 4800\ {\AA}. The median value in each aperture is included in the line labels.}
\label{fig:sn_hist}
\end{figure}

\begin{figure}
\includegraphics[width=\linewidth]{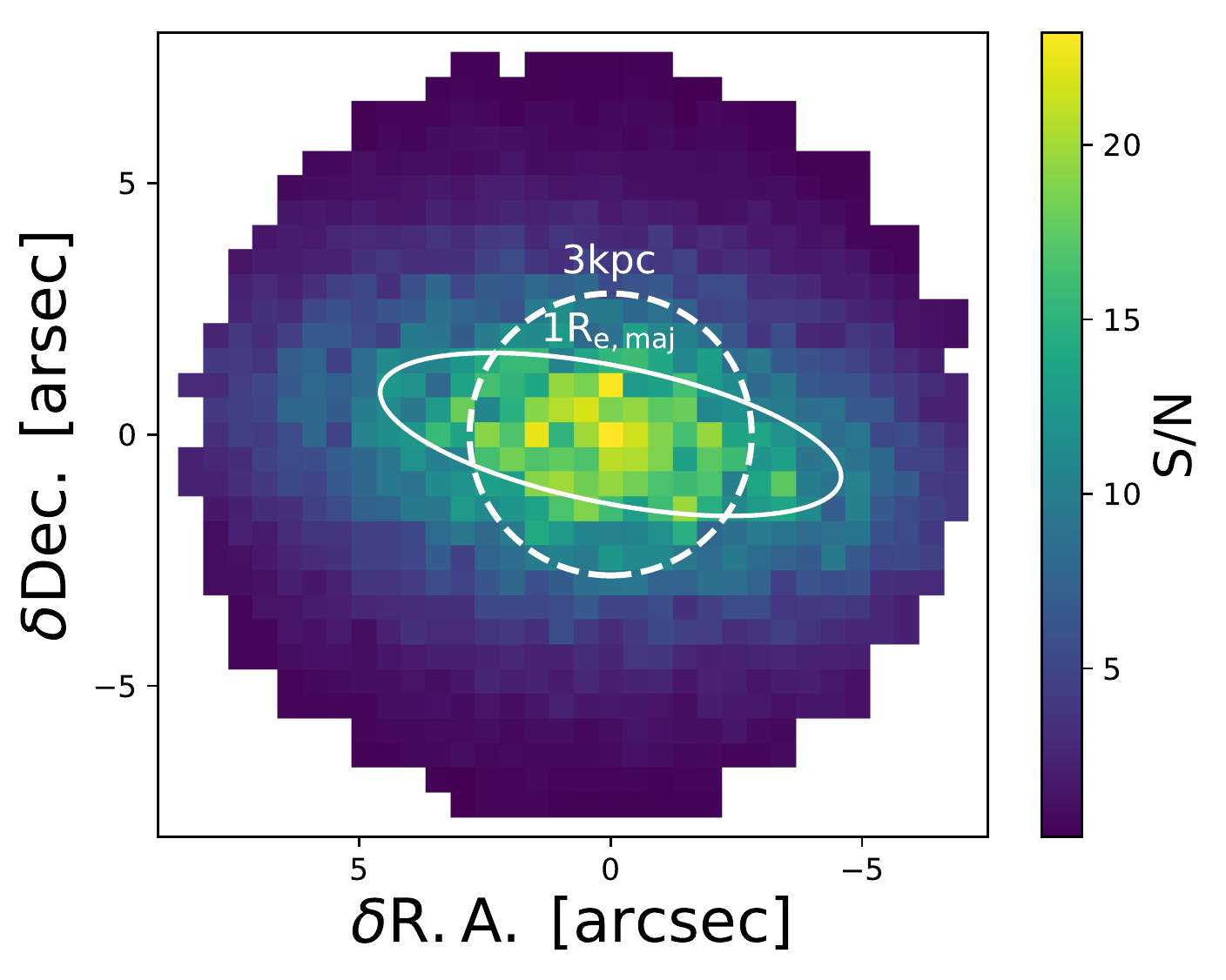}
\caption{The median S/N per {\AA} in the blue cube for GAMA ID 91926. The median is measured between 4600 and 4800\ {\AA}. The white lines show the relative sizes and shapes of two of the six available apertures; the solid line outlines the elliptical aperture with 1 \re\ semi-major axis, and the dashed line shows the circular aperture of 3 kpc diameter. For this galaxy, the S/N per \AA in both the 3 kpc diameter, and \re\ semi-major aperture spectra are 51}
\label{fig:sn_map}
\end{figure}

We estimate the aperture S/N by taking the median S/N value per {\AA} between 4600 and 4800\ {\AA} in the rest-frame. This range is clear of skylines and is fully contained within the SAMI blue arm. In Fig. \ref{fig:sn_hist} we show histograms of the median S/N for the six available apertures. We also show the median S/N map for a typical galaxy in Fig.~\ref{fig:sn_map}, outlining the \re\ and 3kpc apertures for reference. The $1R_e$ aperture spectra have a median S/N of 32; 92\% have a S/N above 10, and 25\% have a value above 50. The median S/N of the central pixel in the data cubes is 14 with a standard deviation of 13.

\section{Value-added data products: Emission-line physics}
\label{sec:emission_line_products}

For each of the core data products listed above (cubes, binned cubes, and aperture spectra), we fit strong emission lines arising from ionised gas and extract line fluxes, velocities, and velocity dispersions. From these we then create maps of value-added products such as extinction maps (derived from the Balmer decrement), star-formation rate surface densities, and excitation mechanism classifications. We summarise here the emission line analysis, but note the methods and products are very similar to those released in DR1. We refer the reader to \citep{green2018}, \citet{Ho2016a} and \citet{Medling2018} for further details on the emission-line fitting and resulting data products.

\subsection{Emission-line fitting}
\label{sec:emission_line_fitting}

We fit seven strong optical emission lines within the SAMI wavelength range: [{\sc oii}]3726+3729, H$\beta$, [{\sc oiii}]5007, [{\sc oi}]6300, H$\alpha$, [{\sc nii}]6583, [{\sc sii}]6716 and [{\sc sii}]6731. We also fit the lines [{\sc oiii}]4959 and [{\sc nii}]6548, but their fluxes are fixed to their physical ratios relative to the stronger [{\sc oiii}] and [{\sc nii}] lines. Using version 1.1 of the {\sc lzifu} software package \citep{ho2016b}, we stitch together the blue and red spectra accounting for the differing spectral resolution (Section \ref{subsec:spectral_resolution}). We then subtract the underlying stellar continuum before fitting each emission line with one to three Gaussian profiles. The Gaussian profiles are fit using the Levenberg-Marquardt least-square method implemented in \citep[{\sc mpfit;}][]{markwardt2009}. All selected emission lines are then fit simultaneously with each kinematic component constrained to the same velocity and velocity dispersion. From the Gaussian fits, we obtain the emission line fluxes, velocities and velocity dispersions.

For the spectral cubes (Section \ref{subsec:cubes}), we follow DR1 in providing both a 1-component Gaussian fit capturing the bulk emission and gas-motions, and a multicomponent fit. The multicomponent fit captures both the dominant gas emission and fainter velocity structures such as outflows, and represents a more accurate total gas emission. Each spaxel is fit 3 times using {\sc lzifu} to obtain one, two and three component fits to each emission line. The number of components in the multicomponent fits are determined using an artificial neural network trained by astronomers \citep[for full details on the neural network, and precision success with SAMI data, see][]{hampton2017}. 

One significant difference between the emission line fits to the spectral cubes provided in DR2 relative to DR1 is the fitting of the underlying stellar continuum. In DR1, the stellar continuum was fit on a spaxel-by-spaxel basis using the
penalized pixel-fitting routine \citep[{\sc pPXF};][]{cappellari2004,cappellari2017}, even when the signal-to-noise in the continuum was low. Doing so can lead to large uncertainties in the correction for the underlying absorption lines, specifically impacting the Balmer emission lines. To account for this impact, we included an additional systematic uncertainty in the Balmer lines \citep[described in][]{Medling2018}. 

In DR2, we now use the significantly improved stellar continuum fitting to better subtract the continuum prior to fitting the emission lines. Here, we give a brief description of the continuum fitting procedure and refer to Owers et al. (in prep.) for further details. We use the Voronoi-binned data, which has S/N$\ \sim10$ in the continuum, to constrain the number of templates that are used to fit each spaxel within the Voronoi bin of interest. This is achieved by using {\sc pPXF} to fit the Voronoi-binned spectrum with a subset of the MILES simple stellar population (SSP) spectral library \citep{vazdekis2010} that contains four metallicities ([M/H] = -0.71, -0.40, 0.00, 0.22) and 13 logarithmically-spaced ages ranging from 0.0063--15.85 Gyrs. Following \citet{cidfernandes2013}, the MILES SSPs are supplemented with younger SSP templates drawn from \citet{gonzalezdelgado2005} with metallicities [M/H] = -0.71, -0.40, 0.00 and ages $0.001-0.025$\,Gyr. During the fitting, emission line templates are included for the Balmer lines, as well as strong forbidden lines. Importantly, this simultaneous fitting of emission and absorption components allows the regions surrounding the age-sensitive Balmer lines to be included in the continuum fits. The stellar kinematics are not fitted for during this process, and are fixed to the values determined in Section~\ref{subsec:stellar_kinematics}. 

The subset of SSP templates that have non-zero weights assigned in the fits to the Voronoi binned spectra are then used during the fitting of each spaxel contained within the region defined by the Vorenoi bin. Again, emission lines are fitted simultaneously, and the stellar kinematics are fixed to those determined in Section~\ref{subsec:stellar_kinematics}, while allowing {\sc pPXF} to re-determine the optimal template weights only for spaxels where the S/N$>5$. For spaxels with S/N$<5$, the weights determined during the Voronoi binned fitting are used to produce a single optimal template, while the stellar kinematics are fixed to those derived from the Voronoi-binned data. This helps to guard against poor fits due to low S/N. In all of the {\sc pPXF} fitting described above, we include a 12th order multiplicative polynomial. This continuum fit is then used in {\sc lzifu} to subtract the continuum and measure the final line fluxes for the 1- to 3-components fits. Overall this method produces similar line fluxes to those found in DR1, but with better constraints in spaxels with low-S/N continua, and some systematic offsets in galaxies with significant Balmer absorption features. 

\begin{figure*}
\centering
\includegraphics[width=\linewidth]{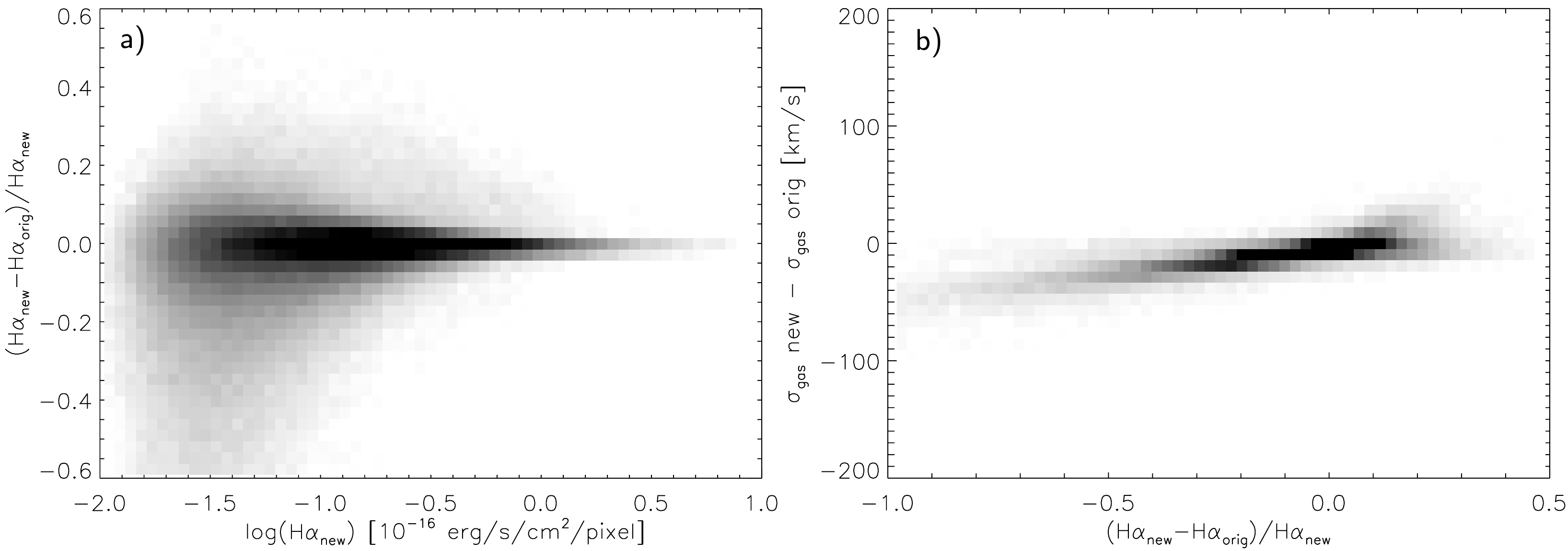}
\caption{Comparison of the LZIFU determined 1-component H$\alpha$ flux based on the original DR1 continuum method and the new DR2 continuum method. All spaxels with weak continuum (S/N$_{\rm cont}<10$) and significant line detection (S/N$_{\rm H\alpha} > 5$) in DR2 are examined here. a) A 2D density histogram of the relative difference in H$\alpha$ flux as a function of the new DR2 continuum H$\alpha$ flux. b) A 2D density histogram showing how the relative difference in H$\alpha$ flux between the two methods correlates with the difference in the determined gas velocity dispersion between the two methods. These figures demonstrate that the majority of spaxels are consistent (>80\%), but at weak line flux ($<2.10^{-17} \rm erg/s/cm^2/pixel$), errors in the Balmer line absorption feature due poor continuum fitting in the original DR1 method lead to weak but incorrect broad lines. }\label{fig:Newcont_comp}
\end{figure*}

\begin{figure*}
\centering
\includegraphics[width=\linewidth,clip,trim = 70 60 60 60]{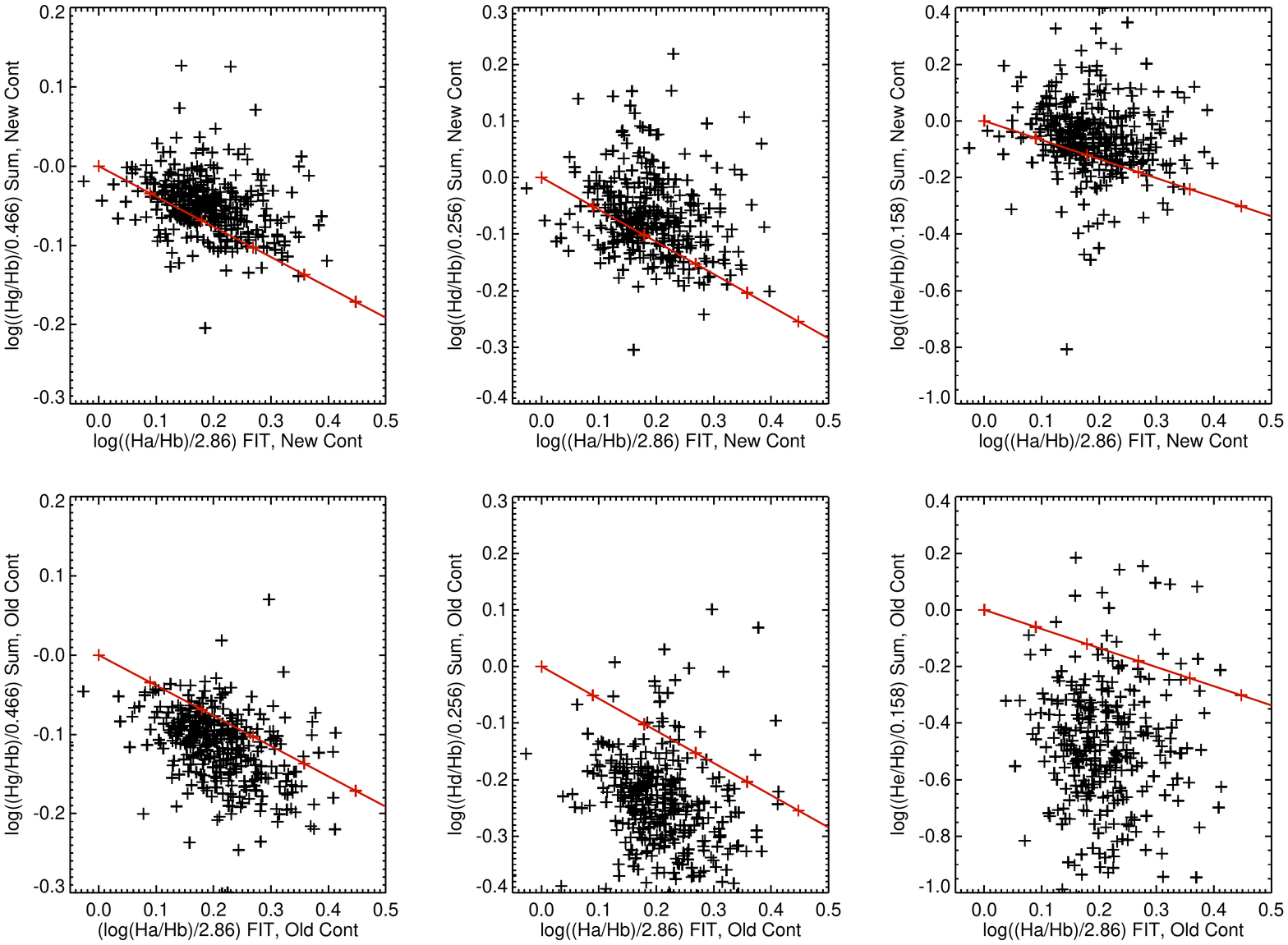} 
\caption{Comparison of the higher-order Balmer decrements derived from emission line fluxes measured after subtracting continua using the continuum fitting method used in DR2 (top row) and the continuum method used in DR1 (bottom row). The comparisons are made for H$\gamma$/H$\beta$, H$\delta$/H$\beta$ and H$\epsilon$/H$\beta$, determined by direct summation over a narrow window around the line centre (left, middle and right panels), with the H$\beta$/H$\alpha$ ratio shown on the x-axis, taken from the {\sc LZIFU} fits. The red line shows the trend in the Balmer decrements of interest derived from a \citet{calzetti2000} extinction law. We see an improvement in both the magnitude of the offset, and the scatter in the distribution when comparing DR2 with DR1, although the offsets from the \citet{calzetti2000} line are in the opposite sense.
\label{fig:BD_comp}}
\end{figure*}

In terms of the H$\alpha$ emission line flux, for most pixels with low continuum S/N the new continuum fitting procedure produces a similar line flux, especially for strong lines. Figure \ref{fig:Newcont_comp}a shows a 2D density diagram of the relative difference in 1-component H$\alpha$ flux for the original DR1 and new DR2 continuum fitting methods for all spaxels with significant line emission (H$\alpha$ S/N $> 5$) and weak continuum (S/N$_{\rm cont}< 10$). The median relative difference for all spaxels is 0.003, but with a standard deviation of 0.3, with these differences clearly increasing for low line fluxes. The majority of spaxels ($\sim 80$\%) have H$\alpha$ fluxes that agree within 10\%, and only $\sim 6$\% of the low continuum S/N spaxels disagree by more than 30\%. 

To reveal the origin of this change we show the relative difference of H$\alpha$ flux against the difference in determined gas velocity dispersion in Figure \ref{fig:Newcont_comp}b. We again show this for all spaxels with significant line emission (H$\alpha$ S/N $> 5$) and weak continuum (S/N$_{\rm cont}< 10$). All spaxels with similar H$\alpha$ fluxes show similar velocity dispersions, but for spaxels with a substantially different H$\alpha$ flux we see that the velocity dispersion of the original method is also systemically different, arising from an improperly determined Balmer absorption feature in the low continuum S/N.

With the aim of understanding the systematic effects of the different continuum fitting procedures between DR1 and DR2, we investigate whether the Balmer decrements measured using higher-order Balmer emission lines depart from the expectations of a Calzetti dust extinction law. These tests are similar to those performed by \citet{groves2012} for the SDSS. To do this, we select a subset of the DR2 galaxies that have more than 10 high-quality (continuum S/N$>5$) spaxels that are classified as star-forming, and have well-detected H$\beta$ and H$\alpha$ fluxes (S/N$>5$). For the high-quality star-forming spaxels in these galaxies, we measure emission fluxes for the Balmer lines H$\epsilon$, H$\delta$, H$\gamma$ and H$\beta$ using direct summation in windows surrounding those lines. The window width is set to $\pm\-3\sigma_{win}$ around the redshifted line centre, where $\sigma_{win}$ is defined by the quadrature sum of the instrumental resolution and the gas-velocity dispersion determined by the 1-component {\sc lzifu} fits. The redshift used to determine the line centre includes contributions from both the galaxy redshift and the gas velocity from {\sc lzifu}. For each line, we measure two sets of emission line fluxes: one after subtracting the continuum defined using the procedure outlined for DR1, and another after subtracting the continuum measured as outlined above. The emission line fluxes are corrected for Galactic extinction, and then used to measure Balmer decrements for the higher order lines. 

In Figure~\ref{fig:BD_comp} for each galaxy we plot the median value of the higher order Balmer decrements against the median H$\alpha$/H$\beta$ decrements, where H$\alpha$ and H$\beta$ are determined from the {\sc lzifu} fits, and are also corrected for Galactic extinction. The decrements are normalised by the theoretical value for Case B recombination. The red-line shows the expected trend due to a \citet{calzetti2000} extinction law. In the top panels, the results derived using the new continuum fitting are shown, while the bottom panels show the results from the DR1 continuum fitting method. Neither the new continuum fitting method used for DR2, nor the method used in DR1 produce results that align with the expectations of a \citet{calzetti2000} extinction law. However, the DR2 results show both a smaller offset and scatter when compared with the DR1 results, and this is particularly true for the H$\delta$/H$\beta$ and H$\epsilon$/H$\beta$ ratios shown in the middle and right-most panels, respectively. 

Given that we have selected only relatively high S/N spaxels, the difference in the results from the two continuum fitting methods is likely driven by two changes. First, in DR2 we now modulate the SSP templates with a multiplicative polynomial rather than the additive polynomial used in DR1. Second, we now simultaneously fit for emission and absorption in the vicinity of the Balmer lines, whereas previously these regions were masked during the fit. The combined effect of using an additive polynomial alongside masking the age-sensitive Balmer lines was that younger templates were often excluded from the fit; the blue flux was modelled by the additive polynomial rather than a young stellar population, thereby underestimating the Balmer absorption at bluer wavelengths. We note that the offset observed in the new DR2 values in Figure~\ref{fig:BD_comp} is similar to that noted by \citet{groves2012} for SDSS DR7 data. Investigation into the origin of the offset is ongoing. At this stage this alternative continuum fitting approach has been applied to the original spectral cubes only because they are more susceptible to inaccurate continuum subtraction due to their typically lower S/N than the other spectral data products. 

In addition to the original spectral cubes, we also provide emission line fits for the binned cubes (Section \ref{subsec:binned_cubes}) and aperture spectra (Section \ref{subsec:aperture_spectra}). For the adaptively-binned and sectors-binned spectral cubes, we follow the same conventions as for the original spectral cubes, providing both 1- and multi-component fits. 

For the annular-binned spectral cubes we provide only 2-component fits, and fit the stellar continuum directly within {\sc lzifu} using {\sc pPXF} given the higher continuum S/N. Only 2-component fits are provided given that in many cases rotation dominates the emission structure in the outer bins leading to double-horned profiles that require two separate components to be fit.

All aperture spectra are also fit using {\sc lzifu}, treating each spectrum as an individual spaxel. As with the original cubes we provide both 1-component and multi-component fits to the aperture spectra. These are provided as tabulated line fluxes, gas velocities (relative to the input heliocentric GAMA redshifts) and velocity dispersions, and associated errors for all apertures described in Section \ref{subsec:aperture_spectra}.

\subsection{Star formation rates and other products}

As in DR1, we also release higher-order data products based upon the emission line fitting; Balmer decrement-based attenuation maps, classification of star-forming regions, and star formation rates. For full details on the determination of these products we refer the reader to \citet{Medling2018}, but briefly summarize these here.

We present the attenuation maps as correction factors, $F_{\rm H\alpha}$ for the H$\alpha$ emission line. Using the Balmer decrement (H$\alpha$/H$\beta$) and assuming a \citet{cardelli1989} extinction law we determine this as;
\begin{equation}
F_{\rm attenuate}=\left(\frac{1}{2.86}\frac{\rm H\alpha}{\rm H\beta}\right)^{2.36},
\end{equation}
where H$\alpha$/H$\beta_{\rm intr}=2.86$ is the intrinsic flux ratio of the Balmer decrement (assuming Case B recombination, $T_{e}=10^4$\,K and $n_{e}=100\,{\rm cm^{-3}}$). For regions where the H$\beta$ line is not detected or H$\alpha$/H$\beta<2.86$ we set $F_{\rm attenuate}=1$ and the associated error $\delta F_{\rm attenuate}=0$. Note that the H$\beta$ non-detection regions may also be high attenuation regions, so this correction factor represents a lower limit in these cases. Also note that for the original resolution and adaptively binned data, the Balmer lines need to be smoothed by a Gaussian kernel of FWHM=1.6 spaxels (0.8\arcsec) to account for the different PSFs before the determination of the Balmer decrement and attenuation correction maps 
as described in \citet{Medling2018}, because of the issue of aliasing arising from differential atmospheric refraction \citep[described in detail in][]{green2018}.   

For all spaxels we also classify whether the emission-line spectrum is dominated by photoionisation by massive stars associated with recent star formation, or by other mechanisms (such as AGN, shocks etc). This is done using cuts on emission line ratio based upon the classification scheme described in \citet{kewley2006}, and fully described in \citet{Medling2018}. For both the original and binned data we present these as star formation masks, where any spaxel dominated by mechanisms other than star formation are set to 0.

We present star formation rate maps for both original and all binned cubes by first creating attenuation-corrected, star formation dominated H$\alpha$ maps. We then convert this to a star formation rate using the \citet{kennicutt1994} calibration converted to a \citet{chabrier2003} stellar initial mass function:
\begin{equation}
{\rm SFR\, [M_{\odot}\, yr^{-1}]=5.16\times10^{-42}}F_{\rm H\alpha}\,[{\rm erg\,s^{-1}}].
\end{equation}
Note that, due to both the removal of contaminated regions via the star formation masks and missing heavily obscured regions where H$\beta$ and even possibly H$\alpha$ are undetected, these maps likely represent lower limits to the true current star formation in the galaxies.

\section{Value-added data products: Absorption-line physics}
\label{sec:absorption_line_products}

\subsection{Stellar kinematics}
\label{subsec:stellar_kinematics}

\subsubsection{Method}
\label{subsubsec:kinematic_method}

Stellar kinematic parameters are extracted from the SAMI cubed data following the method described in detail in \citet{vandesande2017a}. We use the {\sc pPXF} code to fit all spectra. Our method is summarised below. 

SAMI blue and red spectra are combined by first convolving the red spectra to match the instrumental resolution in the blue. We use the code \textsc{log\_rebin} provided with the \textsc{pPXF} package to rebin the combined blue and red spectra onto a logarithmic wavelength scale with constant velocity spacing (57.9 \kms). We use annular binned spectra (Section \ref{subsec:binned_cubes}) to derive local optimal templates from the MILES stellar library \citep{sanchezblazquez2006} that consists of 985 stars spanning a large range in stellar atmospheric parameters. A Gaussian line-of-sight velocity distribution (LOSVD) is assumed, i.e., we extract only the stellar velocity $V$ and stellar velocity dispersion $\sigma$. 

After the optimal template is constructed for each annular bin, we run \textsc{pPXF} three times on each galaxy spaxel. For every step, we mask the following emission lines: [OII], H$\delta$, H$\gamma$, H$\beta$, [OIII], [OI], H$\alpha$, [NII], and [SII], even if no emission lines are detected. The first fit is used for determining a precise measure of the noise scaling from the residual of the fit. We use an additive Legendre polynomial to remove residuals from small errors in the flux calibration, and as in \citet{vandesande2017a} we demonstrated that a 12th order additive Legendre polynomial is sufficient for SAMI data. In the second fit, we clip outliers using the CLEAN parameter in \textsc{pPXF}. In the third and final iteration, \textsc{pPXF} is allowed to use the optimal templates from the annular bin in which the spaxel is located, as well as the optimal templates from neighbouring annular bins.

Uncertainties on the LOSVD parameters are estimated using a Monte-Carlo approach. We estimate the uncertainties on the LOSVD parameters for each spaxel from the residuals of the best fit minus the observed spectrum. These residuals are then randomly rearranged in wavelength and added to the best-fitting template. This simulated spectrum is refitted with \textsc{pPXF}, and we repeat the process 150 times. The uncertainties on the LOSVD parameters are the standard deviations of the resulting simulated distributions.

We follow the same method for the binned data (Section \ref{subsec:binned_cubes}). For the aperture spectra (Section \ref{subsec:aperture_spectra}), we construct an optimal template for each individual aperture and then use the same procedure as described above to extract the LOSVD.

We note that the varying spectral resolution from fibre-to-fibre (Section \ref{subsec:spectral_resolution}) can have a significant impact on the stellar kinematic measurements if the intrinsic stellar dispersion is close to, or less than the instrumental dispersion \citep{federrath2017,zhou2017}. However, the stellar kinematic measurements are obtained from the cubed frames, where a dither pattern is applied. As individual spaxels are constructed from a combination of multiple fibres, the spectral resolution will be an average of all the contributing fibres. We do not consider the changing resolution as a function of wavelength to be significant enough to impact the extracted stellar kinematic measurements, as those are derived from a simultaneous fit over the entire wavelength range, and the full wavelength variation is comparable to the fibre-to-fibre variation.

\subsubsection{S/N Estimate the from Stellar Kinematic Fits}
\label{subsubsec:sn}

\begin{figure*}
\centering
\includegraphics[width=\linewidth]{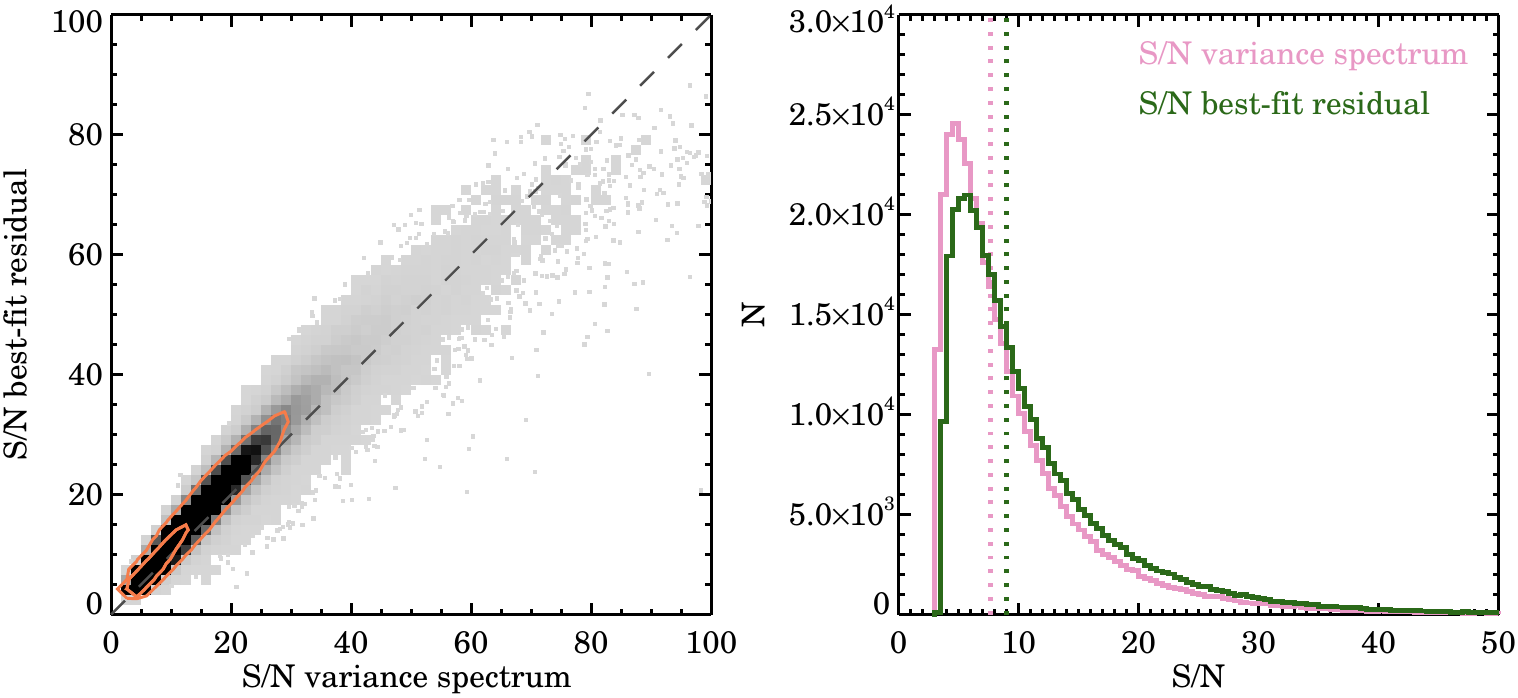} 
\caption{S/N per \AA\ from the flux and variance spectrum compared to the S/N derived from the residual of the stellar kinematic best fit. The grey squares in the left-hand panel show the density of all unbinned spaxels in DR2 that pass the kinematic quality cut (Section \ref{subsubsec:kinematic_quality_cuts}); darker grey means higher density. The orange contours enclose 68 and 95 percent of the data, and the dashed line shows the one-to-one relation. We show the distribution of the two S/N estimates in the right panel. The median S/N from the variance spectrum (pink) is 16.6 percent lower than the median S/N from the best-fit residual (green; 7.70 versus 8.98, respectively).}
\label{fig:sn_resid_cube}
\end{figure*}

Residuals from the observed spectrum minus the best-fitting template provide a good test of the accuracy of the variance spectra. Here, we compare the S/N derived from the flux and variance spectrum to the S/N derived from the stellar kinematic best-fit residual for every unbinned spaxel that meets the quality criteria in the next Section \ref{subsubsec:kinematic_quality_cuts}. The variance spectrum S/N was derived in the blue wavelength region between spectral pixel 1100 and 1600 (approximately between 4500\ \AA\ and 5000\ \AA), whereas the S/N from the residuals is determined for all "goodpixels" in the \textsc{pPXF} fit (i.e., excluding emission lines and $3-\sigma$ outliers). Therefore, we stress that this comparison should be considered a consistency check, not an exact derivation of the cube variance scaling. 

In the left-hand panel of Fig.~\ref{fig:sn_resid_cube} we find that $>95$ percent of the data are above the one-to-one relation, which indicates that the variances may be slightly overestimated. While there are considerably fewer data-points above $S/N \gtrsim 50$, the scatter between the two S/N estimates becomes increasingly larger and the S/N from the best-fit residual drops below the S/N from the variance spectrum. However, we note that at this high S/N, uncertainties from the adopted stellar library, template mismatch, and other fitting related issue are starting to dominate the residual S/N estimate. Thus, above $S/N \sim 50$, the S/N comparison becomes harder to interpret.

We show the distribution of both S/N estimates in the right-hand panel of Fig.~\ref{fig:sn_resid_cube}. Note that we only show data where the S/N estimated from the variance spectrum is greater than $3$\ \AA$^{-1}$, as lower S/N spaxels are not fit by the stellar kinematics pipeline. As before, we find that the median variance spectrum S/N is lower than the S/N derived from the best-fit residual. The difference in medians is 16.6 percent. Thus, while we find that the DR2 variances may be slightly overestimated, we conclude that this offset is relatively mild.

\subsubsection{Kinematic Quality Cuts}
\label{subsubsec:kinematic_quality_cuts}

For SAMI data we recommend applying the following quality criteria to the stellar kinematic data: signal-to-noise (S/N) $>3$\ \AA$^{-1}$, \sobs $>$ FWHM$_{\rm{instr}}/2 \sim 35$\kms, $V_{\rm{error}}<30$\kms\ \citep[Q$_1$ from][]{vandesande2017a}, and $\sigma_{\rm{error}} < \sobs *0.1 + 25$\kms\ \citep[Q$_2$ from][]{vandesande2017a}.

For kinematic data products such as \textsc{kinemetry}, the kinematic position angle, \vs\ and \lr, we have additional flags. We perform a visual inspection of all 1559 SAMI kinematic maps and exclude 42 galaxies with irregular kinematic maps due to nearby objects or mergers that influence the stellar kinematics of the main object. We furthermore exclude 481 galaxies where $\re<1\farcs5$ or where either \re\ or the radius out to which we can accurately measure the stellar kinematics is less than the half-width at half-maximum of the PSF (HWHM$_{\rm{PSF}}$). This brings the final number of galaxies for which we can derive reliable 2D stellar kinematic data products to 1036. The number of galaxies with \re\ aperture velocity dispersion measurements is 1171.

\begin{figure}
\centering
\includegraphics[width=\linewidth]{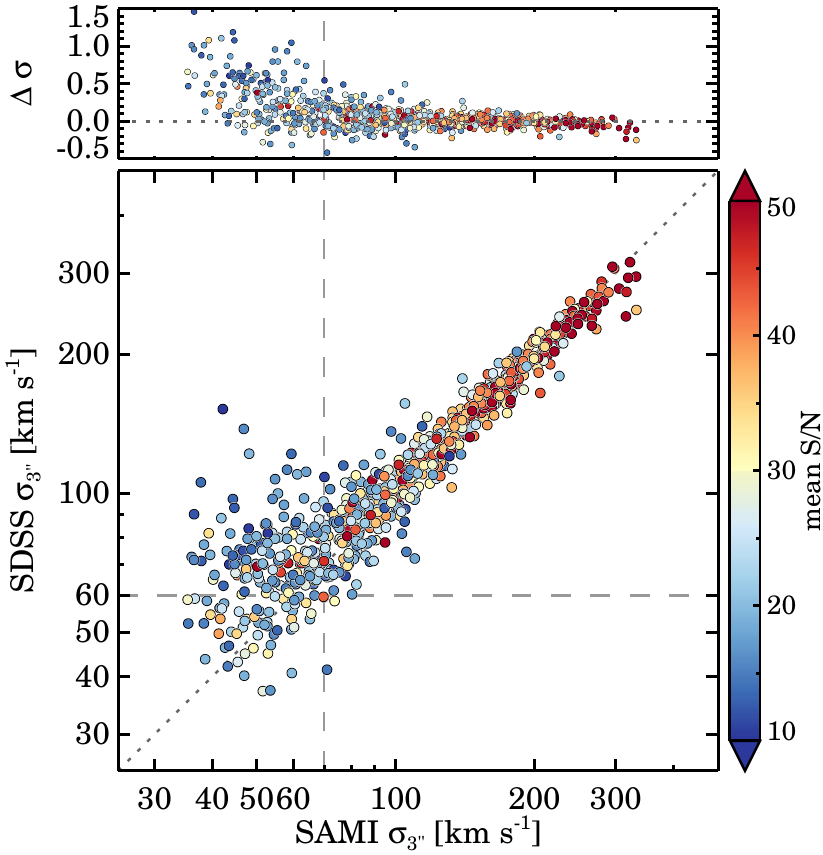} 
\caption{Comparison of the stellar velocity dispersions from the SAMI 3\arcsec\ aperture spectra versus 3\arcsec\ SDSS spectra (main panel). In the top panel we show the ratio of the velocity dispersions, $\Delta \sigma$, which is defined as ($\sigma_{\rm{SDSS}} - \sigma_{\rm{SAMI}} ) / \sigma_{\rm{SAMI}}$. The data are colour coded by the mean continuum S/N of the SDSS single-fibre spectra. The dashed lines show the adopted instrumental resolution of the SAMI blue arm and the SDSS spectrograph, whereas the dotted line shows the one-to-one relation.
Above the spectral resolution limits of both spectrographs, we find an excellent agreement. Below the instrumental resolution the scatter increases due to the lower S/N of these spectra, but there is also an increasing offset from the one-to-one relation.}
\label{fig:sigma_aper_sami_sdss}
\end{figure}

\subsubsection{Aperture Velocity Dispersion Comparison}
\label{subsubsec:sigma_aper_sdss_comp}
We compare our aperture velocity dispersion values to measurements from SDSS 3\arcsec\ single fibre spectra. We refit the SDSS spectra using the same technique and templates as described in Section \ref{subsubsec:kinematic_method}. Our values are in good agreement with the standard SDSS pipeline velocity dispersions and the measurements by \citet{thomas2011}. We show the comparison in Fig.~\ref{fig:sigma_aper_sami_sdss}, where any measurements with (S/N)$_{\rm{SDSS}}<5$\ \AA$^{-1}$, \sobs $<$ FWHM$_{\rm{instr}}/2$, and $\sigma_{\rm{error}} > \sobs *0.025 + 10$ \kms are excluded. The first and last quality criteria are stricter than described in Section \ref{subsubsec:kinematic_quality_cuts} due to the higher mean S/N of the SAMI and SDSS aperture spectra.  We quantify the comparison using the median $\sigma$ offset $(\sigma_{\rm{SDSS}} - \sigma_{\rm{SAMI}})$ and the fractional dispersion difference, $\Delta\sigma = (\sigma_{\rm{SDSS}} - \sigma_{\rm{SAMI}} ) / \sigma_{\rm{SAMI}}$. There is excellent agreement between the SAMI and SDSS measurements when the velocity dispersion is higher than the SAMI instrumental resolution of $\sim70$ \kms\ (median $\sigma$ offset = 1.4 \kms, RMS $\Delta \sigma$ = 0.10), or when the S/N of the SAMI aperture spectra is greater than 25 per \AA\ (median $\sigma$ offset = 1.9 \kms, RMS $\Delta \sigma$ = 0.12). Below 70 \kms, we find a larger median $\sigma$ offset = 14.9 \kms\ and the RMS scatter in $\Delta \sigma$ increases to 0.47, but we note that the majority of galaxies with low velocity dispersion also have the lowest S/N.

Below the instrumental resolution we detect a deviation from the one-to-one relation; the SDSS velocity dispersions are increasingly higher than the SAMI measurements. This hints at a systematic bias in the SDSS velocity dispersions at low $\sigma$. Note that it is extremely difficult to reconstruct an exact SDSS single-fibre measurement using IFS data; the SDSS aperture is circular, convolved with an observed PSF, and differential atmospheric refraction cannot be corrected for. Thus, we believe the SAMI aperture spectra to be more reliable for three reasons. First, the SDSS spectrograph is mounted on the telescope near the Cassegrain focus and therefore suffers from flexure that can change the spectral resolution, whereas the SAMI-AAOmega is a bench-mounted spectrograph and is therefore more stable. Secondly, the S/N in the SAMI spectra is higher, and thirdly our simulations of the recovery of $\sigma$ from synthetic SAMI spectra do not reveal a systematic offset below the instrumental resolution \citep{fogarty2015} in the SAMI measurements.

There are two notable outliers at the highest SAMI velocity dispersion, where the SDSS velocity dispersion is significantly lower than the SAMI value ($\sigma_{\rm{SDSS}}\sim250\kms$ versus $\sigma_{\rm{SAMI}}\sim300\kms$, respectively). A closer investigation of these two objects revealed that one galaxy is a merger remnant with strong emission lines, where a small misplacement of the fibre-centre can cause a large change in the velocity dispersion. The second object consists of two galaxies, one in the foreground and another in the background, with the peak flux of the sources separated by less than 3\arcsec. Similar to the first object, a small offset in position results in a large change in $\sigma$.

\subsubsection{Kinematic Asymmetry}
\label{subsubsec:kinemetry}

We estimate the kinematic asymmetry of the galaxy velocity fields in DR2 following the method outlined in \citet{vandesande2017a}. We assume that the velocity field of a galaxy can be described with a simple cosine law along ellipses. Kinematic deviations from the cosine law can be modelled by using Fourier harmonics. The first order decomposition $k_1$ is equivalent to the rotational velocity, whereas the high-order terms ($k_3$, $k_5$) describe the kinematic anomalies. The kinematic asymmetry can be quantified by using the amplitudes of the Fourier harmonics. Following \citet{krajnovic2011}, the kinematic asymmetry is defined using the amplitudes of the Fourier harmonics ratio $k_5/k_1$.

We determine the amplitude of the Fourier harmonics on all velocity data that pass the quality cut Q1, measured using the \kinemetry\ routine \citep{krajnovic2006,krajnovic2008}. In the fit, the position angle is a free parameter, whereas the ellipticity is restricted to vary between $\pm0.1$ of the photometric ellipticity. For each ellipse, the \kinemetry\ routine determines a best-fitting amplitude for $k_1$, $k_3$, and $k_5$. We use the SAMI flux images to determine the luminosity-weighted average ratio $k_5/k_1$ within one effective radius. The uncertainty on $k_5/k_1$ for each measurement is estimated from Monte Carlo simulations. 

\subsubsection{Kinematic Position Angle}
\label{subsubsec:kinematic_position_angle}

The position angle (PA) of the stellar rotation was measured from the two-dimensional stellar velocity kinematic maps on all spaxels that pass the quality cut Q1. We use the \textsc{fit\_kinematic\_pa} code that is based on the method described in Appendix C of \citet{krajnovic2006}. The kinematic PA was measured with an assumed centre of the map at (25.5,25.5). We checked whether the kinematic PA is sensitive to the centroid choice by performing the kinematic PA fit an additional four times, each with a different centre position: (25,25), (26,26), (25,26), (26,25). We found no systematic difference between the average of these four fits and our default centre.

\subsubsection{\vs\ and \lr}
\label{subsubsec:vsigma_and_lambdar}

For each galaxy, we use the unbinned flux, velocity, and velocity dispersion maps, to derive the ratio of ordered versus random motions \vs\ using the definition from \citet{cappellari2007}:

\begin{flushleft}
\begin{equation}
\left(\frac{V}{\sigma}\right)^2 \equiv \frac{\langle V^2 \rangle}{\langle \sigma^2 \rangle} = \frac{ \sum_{i=0}^{N_{spx}} F_{i}V_{i}^2}{ \sum_{i=0}^{N_{spx}} F_{i}\sigma_i^2}.
\label{eq:vs}
\end{equation}
\end{flushleft}

\noindent The spin parameter proxy \lr\ is derived from the following definition by \citet{emsellem2007}:

\begin{flushleft}
\begin{equation}
\lambda_{R} = \frac{\langle R |V| \rangle }{\langle R \sqrt{V^2+\sigma^2} \rangle } = \frac{ \sum_{i=0}^{N_{spx}} F_{i}R_{i}|V_{i}|}{ \sum_{i=0}^{N_{spx}} F_{i}R_{i}\sqrt{V_i^2+\sigma_i^2}}.
\label{eq:lr}
\end{equation}
\end{flushleft}

\noindent In both equations, the subscript $i$ refers to the $i^{th}$ spaxel within the ellipse, $F_{i}$ is the flux of the $i^{th}$ spaxel, $V_{i}$ is the stellar velocity in \kms, and $\sigma_{i}$ is the velocity dispersion in \kms. For \lr, $R_{i}$ is the semi-major axis of the ellipse on which spaxel $i$ lies, which is different from other surveys that use the circular projected radius to the centre (e.g., ATLAS$^{\rm{3D}}$, \citealt{emsellem2007}). The sum is taken over all spaxels $N_{spx}$ that pass the quality cut Q1 and Q2 within an ellipse with semi-major axis \re\ and axis ratio $b/a$. 

We require a fill factor of 95 per cent of good spaxels within the aperture for producing \re\ measurements. This selection results in 695 galaxies with \vse\ and \lre\ measurements in DR2. For the data where the largest kinematic aperture radius is smaller than the effective radius or where the effective radius is smaller than the seeing disk (238 galaxies), we apply an aperture correction as described in \citet{vandesande2017b}. Including the aperture corrections, 933 galaxies have \vse\ and \lre\ measurements out of the total 1559 galaxies in DR2 (59.8 per cent).

\subsection{Stellar populations}

Stellar population parameters are measured from the aperture spectra described in Section \ref{subsec:aperture_spectra}, using the method described in \citet{scott2017}, which we briefly summarise below. We note that the \re\ aperture measurements presented as part of this release are not identical to those used in \citet{scott2017} as i) a new internal SAMI version of the reduced data has been used and ii) the apertures are elliptical as opposed to circular and use different measurements of \re.

We began by measuring Lick absorption line strength indices for all spectra \citep{faber1973,worthey1994}. We begin by correcting for ionised gas emission and bad pixels by comparing the observed spectra to a set of MILES SSP template spectra that are unaffected by emission, identifying pixels that differ significantly between the observed spectrum and best fitting linear combination of template spectra, and replacing affected pixels with the values of the template spectra. We then broaden each spectrum to the wavelength-dependent Lick spectral resolution. We measured absorption line strengths for a set of 20 Lick indices on the emission-corrected, broadened spectra following the index definitions of \citet{trager1998}. For spectra where the effective resolution is already broader than the Lick resolution we correct for the effect of intrinsic broadening following \citet{schiavon2007}. Uncertainties on all indices are determined by a Monte Carlo reallocation of the residuals and repeating the measurements on 100 realisations of the spectra.

The observed Lick index measurements are converted to Single Stellar Population (SSP) equivalent age, metallicity, [Z/H], and alpha-abundance, [$\alpha$/Fe]. This conversion is done by comparing the observed absorption line strengths to the predictions of the stellar population synthesis models of \citet{schiavon2007} and \citet{thomas2010}. We use a $\chi^2$ minimisation approach with an iterative rejection of discrepant indices to determine the best-matching SSP parameters, as first implemented by \citet{proctor2004}. For the \citet{thomas2010} models we use all 20 measured indices. For the \citet{schiavon2007} models only 16 of the measured indices are predicted by the stellar population synthesis models. For the reasons outlined in \citet{scott2017}, we use SSP equivalent ages from \citet{schiavon2007} models and SSP equivalent [Z/H] and [$\alpha$/Fe] from \citet{thomas2010} models.

\begin{figure*}
\centering
\includegraphics[width=0.95\textwidth]{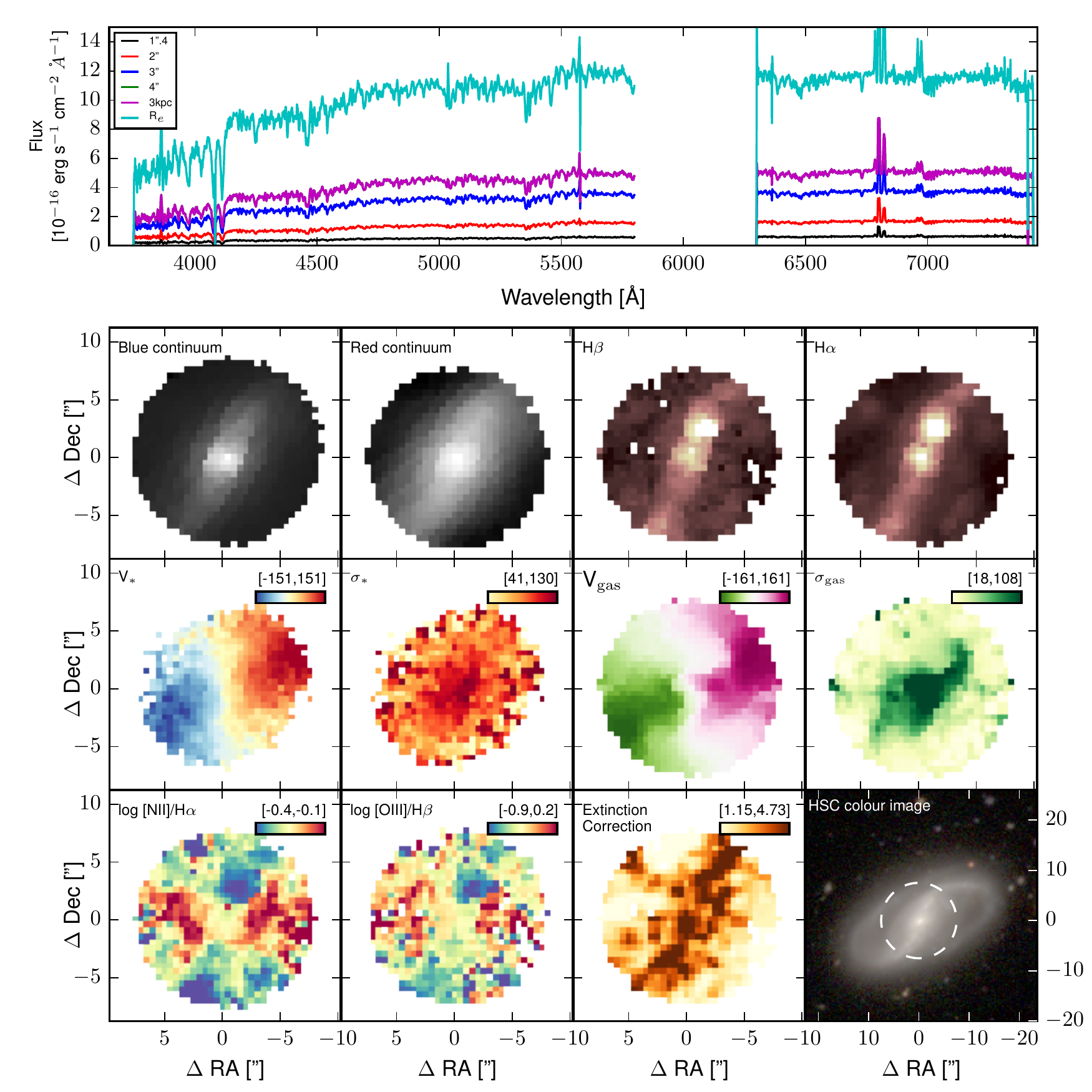}
\caption{Summary of data available for galaxy ID 22887. Along with Table \ref{tab:so_summary}, this figure provides an overview of data available for each galaxy in the release. The upper panel shows the flux-calibrated aperture spectra for the six apertures included in this release. For this object, the 4\arcsec and 3 kpc apertures are identical. The lower panels show, from left to right, top to bottom: log-scaled median flux from the blue cube, log-scaled median flux from the red cube, H$\beta$ emission line flux, H$\alpha$ emission line flux, stellar velocity and stellar velocity dispersion in km s$^{-1}$, gas velocity and velocity dispersion in \kms, log-scaled flux ratio of [NII]6583 to H$\alpha$, log-scaled flux ratio of [OIII]5007 to H$\beta$, Balmer decrement and log-scaled $g'r'i'$ Hyper Suprime-Cam image \citep{aihara2018}. For the upper row, lighter colours indicate higher fluxes. For the lower two rows, the range of the colour scale is indicated in the upper right of each panel, with the sense indicated by the accompanying colour bar. The dashed white circle indicates the location of the SAMI bundle footprint in the HSC image.}
\label{fig:so_summary}
\end{figure*}

\begin{table*}
\centering
\caption{Summary of data available for galaxy ID 22887. Along with Figure \ref{fig:so_summary}, this table provides an overview of data available for each galaxy in the release. Note that this table presents data for only one aperture, the R$_e$ aperture, of the six apertures for which data is available in the release. (1) SAMI/GAMA galaxy ID, (2) Right ascension, (3) Declination, (4) Flow-corrected redshift, (5) Stellar mass, (6) GAMA $r$-band effective radius, (7) GAMA light-weighted $r$-band ellipticity, (8) Mean stellar age, (9) Mean stellar metallicity, (10) Mean stellar $\alpha$-abundance, (11) Stellar velocity dispersion, (12) Ratio of stellar velocity to velocity dispersion, (13) Stellar spin parameter proxy, (14) Photometric position angle, (15) Stellar kinematic position angle, (16) Gas velocity dispersion, (17) Star formation rate, (18) H$\alpha$ flux, (19) H$\beta$ flux, (20) Ratio of N[II] to H$\alpha$ flux, (21) Ratio of [OIII] to H$\beta$ flux.}
\label{tab:so_summary}
\begin{tabular}{l c c c c c c c c c c c c}
\hline
\hline
CATID & RA(J2000) & Dec(J2000) & $z$ & $\log \mstar$ & \re & $\epsilon$ & Age & [Z/H] & [$\alpha$/Fe] & $\sigma_{\rm{e,stars}}$ \\
\hline
 & deg & deg & & M$_\odot$ & arcsec & & Gyrs & & & \kms \\
 (1) & (2) & (3) & (4) & (5) & (6) & (7) & (8) & (9) & (10) & (11) \\
\hline
22887 & 179.403 & 1.14077 & 0.0375 & 10.47 & 6.2 & 0.54 & $2.1\pm0.5$ & $-0.08\pm0.14$ & $0.04\pm0.11$ & $124\pm1$ \\
\hline
\end{tabular}
\begin{tabular}{c c c c c c c c c c}
\hline
($V/\sigma$)$_{\rm{e,stars}}$ &$\lambda_\mathrm{R}$ & PA$_\mathrm{phot}$ & PA$_\mathrm{stars}$ & $\sigma_{\rm{e,gas}}$ & SFR & H$_\alpha$ flux & H$_\beta$ flux & log([NII]/H$_\alpha$) & log([OIII]/H$_\beta$)\\
\hline
& & deg & deg & \kms & M$_\odot$ yr$^{-1}$ & 10$ ^{-16}$ ergs & 10$ ^{-16}$ ergs & & \\
& & & & & & cm$^{-2}$ s$^{-1}$ & cm$^{-2}$ s$^{-1}$ & & \\
(12) & (13) & (14) & (15) & (16) & (17) & (18) & (19) & (20) & (21) \\
\hline
$0.62\pm0.01$ & $0.51 \pm 0.01$ & -36 & 108 & 101 & 0.72 & $116.7\pm0.3$ & $22.9\pm0.3$ & -0.34 & -0.49\\
\hline
\hline
\end{tabular}
\end{table*}

\begin{figure*}
\centering
\includegraphics[width=0.95\textwidth]{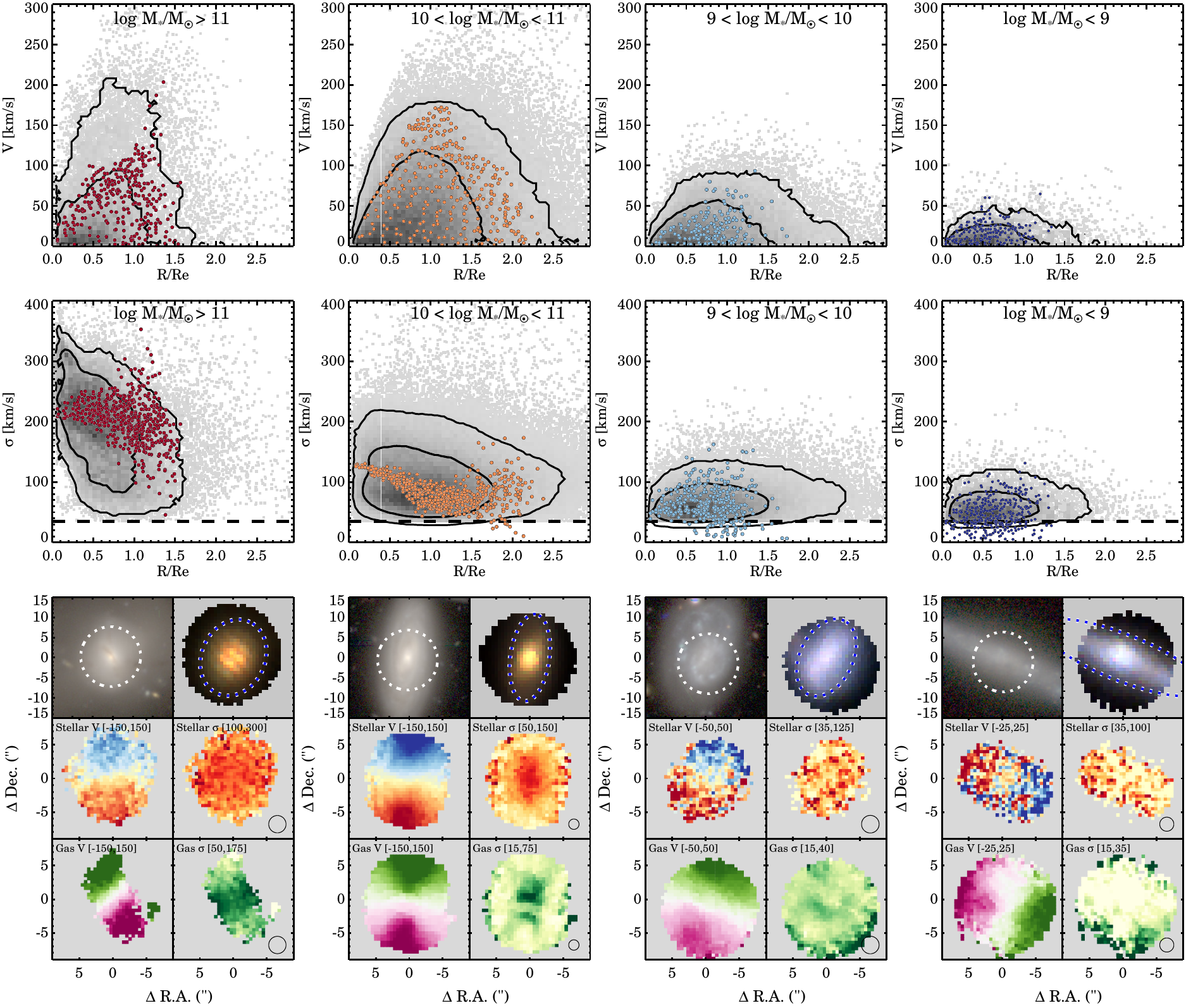}
\caption{Distribution of individual spaxel velocity (top-row) and velocity dispersions (2$^{\rm{nd}}$-row) as a function of radius, divided into four stellar mass bins. The contours show 68\% and 95\% of the data, and are smoothed with a boxcar filter. The dashed line indicates the adopted velocity dispersion limit of 35 \kms\ as described in Section \ref{subsubsec:kinematic_quality_cuts}. We also show a Hyper Suprime-Cam $g'r'i'$ combined colour image, a SAMI $g^sr^si^s$ colour image reconstructed from the spectra, and the stellar and gas velocity $V$ and velocity dispersion $\sigma$ maps. We indicate the range of the kinematic maps between ``[ ]" brackets. For the velocity fields, blue (stellar) and green (gas) indicate negative velocities, whereas red (stellar) and purple (gas) show positive velocities.
For the velocity dispersion maps, yellow-to-red and beige-to-green indicate low-to-high values.
In the HSC colour image, the white circle indicates the SAMI hexabundle field-of-view. The blue dashed ellipse in the SAMI reconstructed colour image indicates one effective radius. The small circle in the stellar and gas velocity dispersion map shows the size of the PSF-FWHM.
For each mass bin, we highlight one galaxy with coloured points and the image and maps; from left to right we show GAMA560883, GAMA84900, GAMA8353 and GAMA16863. We refer to the text for details on each individual galaxy.}
\label{fig:radial_d_sigma_examples}
\end{figure*}
%

In this release we make available all 20 Lick index measurements and the three SSP equivalent stellar population measurements for all 6 apertures. For the Lick index measurements, we provide a flag for each aperture to indicate where the measurements may be unreliable, predominantly due to low S/N. For the SSP equivalent measurements, we provide age and [Z/H] only where the S/N of the given aperture spectrum is greater than 10 per \AA. We provide [$\alpha$/Fe] only where the S/N of the aperture spectrum is greater than 20 per \AA. 

\section{Online Data}
\label{sec:data_access}

In Figure \ref{fig:so_summary} and Table \ref{tab:so_summary} we provide an overview of some of the data available for each of the 1559 galaxies in DR2. Figure \ref{fig:so_summary} presents the spatially resolved measurements, while aperture measurements and other tabular data are presented in Table \ref{tab:so_summary}. In the following subsections we describe how these data can be accessed, searched for and downloaded, and provide an illustration of how the combined data products can be used.

\subsection{Data access}

\begin{table}
\centering
\caption{Summary of three dimensional data products included in DR2}
\label{tab:3d_products}
\begin{tabular}{l r}
\hline
Product & Versions \\
\hline
Default Cubes & Blue/Red\\
Adaptive Binned Cubes & Blue/Red\\
Annular binned Cubes & Blue/Red\\
Sectors Binned Cubes & Blue/Red\\
\hline
\end{tabular}
\end{table}

\begin{table}
\centering
\caption{Summary of two dimensional data products included in DR2. 1-comp, 2-comp and recom-comp refer to the one component, two component and recommended component {\sc LZIFU} fits, see Section \ref{sec:emission_line_fitting} for details. For the annular binning scheme we provide only two component data. For the other binning schemes and the unbinned data we provide both one component and recommended component data.}
\label{tab:2d_products}
\begin{tabular}{l r}
\hline
Product & Versions\\
\hline
Stellar Velocity & n/a\\
Stellar Velocity Dispersion & n/a\\
Gas Velocity & 1-comp, 2-comp, recom-comp\\
Gas Velocity Dispersion & 1-comp, 2-comp, recom-comp\\
Star Formation Rate Density & 1-comp, 2-comp, recom-comp\\
Balmer Decrement & 1-comp, 2-comp, recom-comp\\
H$\alpha$ flux & 1-comp, 2-comp, recom-comp\\
H$\beta$ flux & 1-comp, 2-comp, recom-comp\\\relax
[{\sc oii}]3726+3729 flux & 1-comp, 2-comp, recom-comp\\\relax
[{\sc oiii}]5007 flux & 1-comp, 2-comp, recom-comp\\\relax
[{\sc oi}]6300 flux & 1-comp, 2-comp, recom-comp\\\relax
[{\sc nii}]6583 flux & 1-comp, 2-comp, recom-comp\\\relax
[{\sc sii}]6716 flux & 1-comp, 2-comp, recom-comp\\\relax
[{\sc sii}]6731 flux & 1-comp, 2-comp, recom-comp\\
\hline
\end{tabular}
\end{table}

\begin{table}
\centering
\caption{Summary of one dimensional data products included in DR2}
\label{tab:1d_products}
\begin{tabular}{l r}
\hline
Product & Versions \\
\hline
3kpc circular aperture spectrum & Blue/Red \\
\re\ elliptical aperture spectrum & Blue/Red \\
1\farcs4 circular aperture spectrum & Blue/Red \\
2\arcsec circular aperture spectrum & Blue/Red \\
3\arcsec circular aperture spectrum & Blue/Red \\
4\arcsec circular aperture spectrum & Blue/Red \\
\hline
\end{tabular}
\end{table}

\begin{table}
\centering
\caption{Summary of data tables included in DR2. All tables are fully queryable through the Data Central interface.}
\label{tab:catalogues}
\begin{tabular}{p{2.5cm} p{5cm}}
\hline
Catalogue & Summary \\
\hline
Sample & General galaxy properties including \mstar, \re, $z$, morphological classification etc.\\
Lick Indices & Measurements and uncertainties of Lick absorption line strengths from aperture spectra.\\
SSP Values & Measurements and uncertainties of SSP-equivalent age, metallicity and [$\alpha$/Fe] from aperture spectra.\\
Aperture Stellar Kinematics & Measurements and uncertainties of stellar kinematic quantities from aperture spectra.\\
Resolved Stellar Kinematics & Measurements and uncertainties of stellar kinematic quantities from maps.\\
Aperture Emission Line Properties & Measurements and uncertainties of emission line quantities from aperture spectra.\\
\hline
\end{tabular}
\end{table}

The data of this release are available through an online database provided by Australian Astronomical Optics' Data Central\footnote{https://datacentral.org.au/} service. Data Central delivers a variety of astronomical data sets of significance to Australian astronomical research. As well as storing and serving data, Data Central provides a flexible query tool, allowing users to search for and link data across multiple surveys. Users can search based not only on source position, but also on any measured property stored in Data Central data tables, allowing highly flexible science queries to be executed.

DR2 data products are available for download as multi-extension {\sc fits} files, while table data are available in a variety of formats including {\sc fits, csv and VOTable} files. The contents and structure of all SAMI data products are fully described in the Data Central schema browser. Further documentation describing how data products are derived, as well as many other details relating to the SAMI Galaxy Survey and DR2 in particular, can be found in Data Central's accompanying documentation. The data products available in DR2 are summarised in Tables \ref{tab:3d_products} to \ref{tab:catalogues}. All table data is searchable. All one-, two- and three-dimensional data products and tables can be downloaded through Data Central.

\subsection{Data Demonstration} 

\begin{figure*}
\centering
\includegraphics[width=\textwidth]{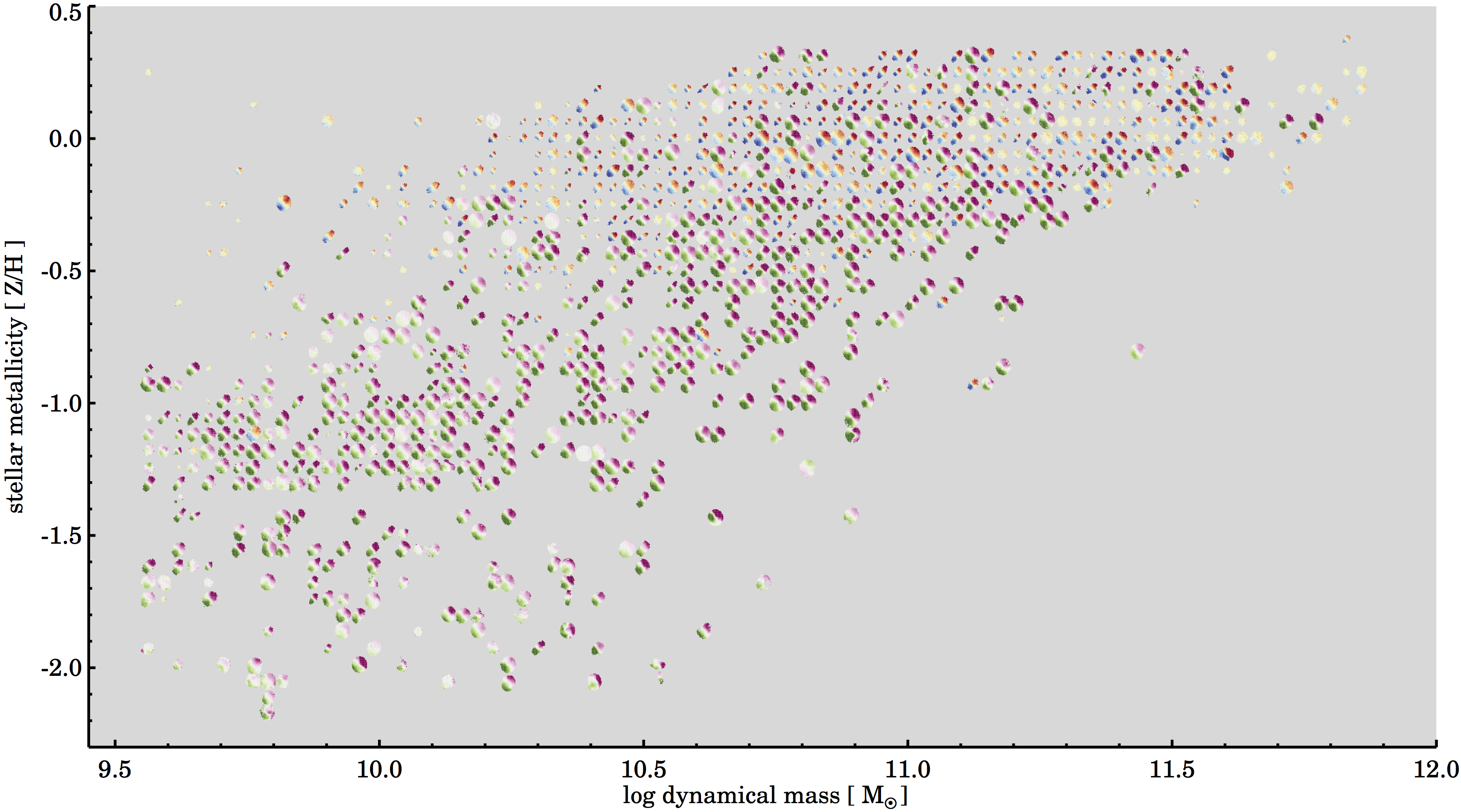}
\caption{Stellar metallicity within one \re\ versus dynamical mass. We show stellar velocity maps for all early-type galaxies (blue-red) and gas velocity maps for the late-type galaxies (green-purple). For each galaxy we show its velocity map aligned to 45$^{\circ}$ using the stellar or gas kinematic position angle, with the velocity range set by the stellar mass Tully-Fisher relation \citep{dutton2011}. A regularization algorithm is applied to avoid overlap of the velocity maps; dynamical masses and metallicities are indicative, not exact. The velocity maps are shown at full resolution; individual galaxies are best viewed in high-zoom in the online version.}
\label{fig:poster_img}
\end{figure*}

In this section, we present a demonstration of the science that can be done with the data products within DR2. In Fig.~\ref{fig:radial_d_sigma_examples}, we show the distribution of the velocity $V$ and velocity dispersion $\sigma$ of \emph{all} individual spaxels as a function of $R/\re$ (normalised elliptical distance from the galaxy centre). Here, we use spaxels with the stellar kinematic quality criteria from Section \ref{subsubsec:kinematic_quality_cuts}. These criteria result in a total of 344,965 spaxels that we divide into four stellar mass bins.

From the lowest to highest mass bin (right--to--left), we find an increasingly high maximum velocity and velocity dispersion. These correlations are the well-known \citet{tully1977} and \citet{faber1976} relations. The declining average rotation as a function of radius is a selection effect, rather than a physical effect. Due to SAMI's round hexabundles any galaxy with an ellipticity  greater than zero, will have more velocity measurements extracted from spaxels along the minor axis than major axis. As the plane of rotation is usually along the major axis, measurements along the minor axis will have relatively low velocity, which causes an apparent decrease at $R/\re>1.5$.

In the highest mass bin (left panel, $\log(\mstar/\msun)>11$), the maximum of the 1-$\sigma$ contour in velocity is lower than for the intermediate mass bin ($10 < \log(\mstar/\msun)<11$), along with an increase in the velocity dispersion. Above $\log(\mstar/\msun)\sim11$ is also the mass regime where most galaxies appear to change from regular rotating oblate spheroids to mildly triaxial, dispersion dominated spheroids \citep[e.g.,][]{cappellari2016}.

For galaxies with $\log(\mstar/\msun)>10$ we find an increase in the velocity dispersion towards the centre due to a dispersion dominated central component or classical bulge. Below $\log(\mstar/\msun)<10$ we find no evidence for an increase in the velocity dispersion near the centre; this was also noticed by \citet{falconbarroso2017}. However, the results from \citet{falconbarroso2017} suggest lower velocity dispersions in the centre as compared to the galaxy outskirts. While our data also show an average radial increase in $\sigma$, this increase is mostly caused by an increase in the measurement uncertainties at large radius. The stellar mass ($\log(\mstar/\msun)\sim10$) at which galaxies transition from having classical bulges \citep[e.g.,][]{kormendy2004}, to a pseudo-bulge was also seen in \citet{fisher2011}, based on photometric analysis of galaxy surface brightness profiles.  

We highlight the stellar velocity and velocity dispersion of four example galaxies using the coloured points. For each example galaxy we also show the Hyper Suprime-Cam \citep[HSC;][]{aihara2018} Subaru Strategic Program {\it gri} combined colour image, a SAMI {\it gri} colour image reconstructed from the spectra, and the stellar and gas $V$ and $\sigma$ maps. Even though we consider any stellar velocity dispersion measurement that is lower than half the instrumental resolution ($\sim35 \kms$) to be unreliable, we nonetheless show these data for individual galaxies in the top row of Figure~\ref{fig:radial_d_sigma_examples} and the velocity dispersion maps. We do this to demonstrate at what stellar mass a large fraction of the spaxels become unusable. 

The example galaxy with the highest stellar mass (first column, GAMA560883, $\log(\mstar/\msun)=11.21$) shows an offset between the photometric and stellar velocity position angle, and between the stellar and gas kinematic position angle. The gas velocity map seems to be aligned with a faint dust lane that is visible in the HSC image when zoomed-in. GAMA84900 (second column) is a typical fast-rotating early-spiral with $\log(\mstar/\msun)=10.44$, where the photometric, stellar velocity, and gas velocity PAs are all perfectly aligned. The central bulge clearly stands out in the stellar velocity dispersion, whereas the gas velocity dispersion shows a complex central structure. The two low-mass late-spiral-type galaxies (third and fourth column), GAMA8353 with $\log(\mstar/\msun)=9.44$, and GAMA16863 with $\log(\mstar/\msun)=8.67$, show no sign of a dispersion dominated bulge. It is evident that this is not caused by the limiting spectral resolution because the majority of data are well above the adopted spectral resolution limit.

Finally, in Fig.~\ref{fig:poster_img} we present an overview of stellar and gas velocity maps in the stellar metallicity versus dynamical mass plane. We show a total of 903 galaxies that have coverage out to one \re\ with a stellar metallicity and stellar velocity dispersion measurement. Dynamical masses are estimated from the circularised effective radius and velocity dispersion measurements using the following expression:
\begin{equation} 
M_{\rm dyn}=\frac{\beta(n)~ \sigma_{\rm e}^2~R_{\rm e, circ} }{G}.
\label{eq:mdyn}
\end{equation}
Here $\beta(n)$ is an analytic expression as a function of the GAMA S\'ersic index, as described by \citet{cappellari2006}:
\begin{equation} 
\beta(n) = 8.87 - 0.831n + 0.0241n^2.
\label{eq:kn}
\end{equation}
Note that this relation has been calibrated using early-type only samples. There is a caveat that the dynamical mass estimates will be more uncertain for late-type than early-type galaxies due to the larger rotational component in late-type galaxies. Instead of using aperture velocity dispersion, dynamical masses can also be estimated from a combination of the 2D velocity and velocity dispersion maps, which improves their accuracy \citep[e.g.,][]{cortese2014,aquino_ort2018}. As the continuum S/N of the aperture spectra is significantly higher than of individual spaxels, here we chose to use dynamical masses derived from aperture spectra to maximise the sample size.

For galaxies visually classified as early-types we show the stellar velocity map using the blue-red colour scheme, whereas for late-types we show the gas velocity maps in green-purple. From Fig.~\ref{fig:poster_img}, we can see that most early-type galaxies are on a high stellar metallicity sequence, whereas late-type galaxies have a large range in metallicity, but typically lie below the early-type galaxies. With increasing dynamical mass, we start to find an increasing amount of galaxies with little rotation.

\section{Summary}
\label{sec:summary}

In this paper, we present the SAMI Galaxy Survey's second data release that includes  data for 1559 galaxies. This release contains galaxies with redshifts between $0.004 < z < 0.113$ and mass range $7.5 < \log(\mstar/\msun) < 11.6$. We release both the primary spectral cubes covering the blue and red optical wavelength ranges, combined with emission and absorption line value-added products. The data are presented online through Australian Astronomical Optics' Data Central.

Observations for the SAMI Galaxy Survey finished May 2018. The next and final data release of the SAMI Galaxy Survey is planned for mid 2020, and will include further data and value-added products. 

\section*{Acknowledgements}

The SAMI Galaxy Survey is based on observations made at the Anglo-Australian Telescope. The Sydney-AAO Multi-object Integral field spectrograph (SAMI) was developed jointly by the University of Sydney and the Australian Astronomical Observatory. The SAMI input catalogue is based on data taken from the Sloan Digital Sky Survey, the GAMA Survey and the VST ATLAS Survey. The SAMI Galaxy Survey is supported by the Australian Research Council Centre of Excellence for All Sky Astrophysics in 3 Dimensions (ASTRO 3D), through project number CE170100013, the Australian Research Council Centre of Excellence for All-sky Astrophysics (CAASTRO), through project number CE110001020, and other participating institutions. The SAMI Galaxy Survey website is http://sami-survey.org/.

NS acknowledges support of a University of Sydney Postdoctoral Research Fellowship. JvdS is funded under Bland-Hawthorn's ARC Laureate Fellowship (FL140100278). SMC acknowledges the support of an Australian Research Council Future Fellowship (FT100100457). BG is the recipient of an Australian Research Council Future Fellowship (FT140101202). MSO acknowledges the funding support from the Australian Research Council through a Future Fellowship (FT140100255). Support for AMM is provided by NASA through Hubble Fellowship grant \#HST-HF2-51377 awarded by the Space Telescope Science Institute, which is operated by the Association of Universities for Research in Astronomy, Inc., for NASA, under contract NAS5-26555. CFe gratefully acknowledges funding provided by the Australian Research Council's Discovery Projects (grants DP150104329 and DP170100603). SB acknowledges the funding support from the Australian Research Council through a Future Fellowship (FT140101166). TMB is supported by an Australian Government Research Training Program Scholarship. MLPG acknowledges the funding received from the European Union's Horizon 2020 research and innovation programme under the Marie Sklodowska-Curie grant agreement No 707693.

{\it Author contributions.} NS and JvdS oversaw DR2, edited the paper and unless explicitly noted wrote the text. NS prepared the core and stellar population data products. JvdS prepared the stellar kinematics data products. BG oversaw preparation of the emission line data products, with assistance from MSO, TP and AMM. MSO produced the continuum fits outlined in Secton~\ref{sec:emission_line_fitting}, as well as Figure~\ref{fig:BD_comp}, and wrote the parts of Section~\ref{sec:emission_line_fitting} describing those products.
LC prepared the visual morphology classifications. SMC is the survey's Principal Investigator. JB oversaw target selection and wrote parts of Section \ref{sec:sample}. SMC wrote Sections \ref{subsec:sky_subtraction_accuracy} and \ref{sec:fluxcal} and provided Figures \ref{fig:skysub} and \ref{fig:fluxcal}. FDE and SO provided the data for Section \ref{sec:wcs} and Figure \ref{fig:wcs}. TMB and CFo wrote Section \ref{sec:ap_sn} and provided Figures \ref{fig:sn_hist} and \ref{fig:sn_map}. BG wrote Section \ref{sec:emission_line_fitting}. DB provided the data for Section \ref{subsubsec:sigma_aper_sdss_comp} and Figure 
\ref{fig:sigma_aper_sami_sdss}. Remaining authors contributed to overall team operations including target catalogue and observing preparation, instrument maintenance, observing at the telescope, writing data reduction and analysis software, managing various pieces of team infrastructure such as the website and data storage systems, and innumerable other tasks critical to the preparation and presentation of a large data set such as this DR2.




\bibliographystyle{mnras}
\bibliography{sami_dr2.bib}
~\\
{\it \footnotesize \noindent$^{1}$Sydney Institute for Astronomy, School of Physics, A28, The University of Sydney, NSW, 2006, Australia\\
$^{2}$ARC Centre of Excellence for All-Sky Astrophysics (CAASTRO)\\
$^{3}$ARC Centre of Excellence for All Sky Astrophysics in 3 Dimensions (ASTRO 3D)\\
$^{4}$Research School for Astronomy \& Astrophysics Australian National University Canberra, ACT 2611, Australia\\
$^{5}$Department of Physics and Astronomy, Macquarie University, NSW 2109, Australia\\
$^{6}$Australian Astronomical Observatory, 105 Delhi Rd, North Ryde, NSW 2113, Australia\\
$^{7}$Cahill Center for Astronomy and Astrophysics California Institute of Technology, MS 249-17 Pasadena, CA 91125, USA\\
$^{8}$School of Physics, University of New South Wales, NSW 2052, Australia\\
$^{9}$AAO-USydney, School of Physics, University of Sydney, NSW 2006, Australia\\
$^{10}$International Centre for Radio Astronomy Research, University of Western Australia, 35 Stirling Highway, Crawley WA 6009, Australia\\
$^{11}$School of Mathematics and Physics, University of Queensland, Brisbane, QLD 4072, Australia\\
$^{12}$Australian Astronomical Optics, Faculty of Science and Engineering, Macquarie University, 105 Delhi Rd, North Ryde, NSW 2113, Australia\\
$^{13}$Leiden Observatory, Leiden University, P.O. Box 9513, NL-2300 RA Leiden, The Netherlands\\
$^{14}$SOFIA Operations Center, USRA, NASA Armstrong Flight Research Center, 2825 East Avenue P, Palmdale, CA 93550, USA\\
$^{15}$Instituto de Astronom\'{i}a, Universidad Nacional Aut\'{o}noma de M\'{e}xico, A. P. 70-264, C.P. 04510 M\'{e}xico, D.F. Mexico\\
$^{16}$Department of Astronomy, University of Wisconsin, 475 North Charter Street, Madison, WI, 53706, USA\\
$^{17}$Centre for Astrophysics and Supercomputing, Swinburne University of Technology, PO Box 218, Hawthorn, VIC 3122, Australia}




\bsp	
\label{lastpage}
\end{document}